\documentclass[sigconf, nonacm]{acmart}
\newcommand\vldbdoi{XX.XX/XXX.XX}

\newcommand\vldbvolume{15}
\newcommand\vldbissue{1}

\newcommand\vldbavailabilityurl{https://github.com/BU-DiSC/endure}
\newcommand\vldbpagestyle{plain} 

\usepackage[utf8]{inputenc}
\usepackage[T1]{fontenc}
\usepackage{hyperref}
\usepackage{url}
\usepackage{booktabs}
\usepackage{multicol}
\usepackage{amsfonts}
\usepackage{amsmath}
\usepackage{enumitem}
\usepackage{nicefrac}
\usepackage{microtype}
\usepackage{moresize}
\usepackage{graphicx}
\usepackage{balance}
\usepackage{subcaption}
\usepackage[show]{chato-notes}

\usepackage[noend, linesnumbered, ruled]{algorithm2e}
\SetKwIF{If}{ElseIf}{Else}{if}{}{else if}{else}{end if}%
\SetKwFor{While}{while}{}{end while}%
\SetKwFor{For}{for}{}{end for}
\SetKwRepeat{Do}{do}{while}

\usepackage{amsmath,amsfonts,bm}

\newtheorem{problem}{Problem}
\newtheorem{definition}{Definition}
\newcommand{\Endure}{{\sc Endure}}
\newcommand{\cost}{{{\ensuremath{\text{C}}}}}
\newcommand{\costvec}{{{\ensuremath{\bf{c}}}}}

\newcommand{\configuration}{{{\ensuremath{\Phi}}}}

\newcommand{\sizeratio}{{{\ensuremath{T}}}}
\newcommand{\mbuf}{{{\ensuremath{m_{\mathrm{buf}}}}}}
\newcommand{\mfilt}{{{\ensuremath{m_{\mathrm{filt}}}}}}
\newcommand{\mtot}{{{\ensuremath{m}}}}
\newcommand{\policy}{{{\ensuremath{\pi}}}}

\newcommand{\asym}{{{\ensuremath{A_\mathrm{rw}}}}}
\newcommand{\querySel}{{{\ensuremath{S_\mathrm{RQ}}}}}

\newcommand{\workload}{{{\ensuremath{\bf{w}}}}}

\newcommand{\columnvec}{{{\ensuremath{\bf{e}}}}}

\newcommand{\obsworkload}{{{\ensuremath{\hat{\bf{w}}}}}}

\newcommand{\nonemptylookup}{{{\ensuremath{z_1}}}}
\newcommand{\emptylookup}{{{\ensuremath{z_0}}}}
\newcommand{\range}{{{\ensuremath{q}}}}
\newcommand{\update}{{{\ensuremath{w}}}}
\newcommand{\benchmark}{{{\ensuremath{\mathcal{B}}}}}

\DeclareMathOperator*{\argmin}{arg\,min}

\newcommand{\nominal}{{\sc Nominal Tuning}}
\newcommand{\robustw}{{\sc Robust Tuning}}

\newcommand{\etal}{\emph{et al.}}

\newcommand\Paragraph[1]{\vspace{0.02in}  \noindent \textbf{#1.}}

\newcommand{\squishlist}{\begin{list}{$\bullet$}
  { \setlength{\itemsep}{0pt}
     \setlength{\parsep}{3pt}
     \setlength{\topsep}{3pt}
     \setlength{\partopsep}{0pt}
     \setlength{\leftmargin}{1.5em}
     \setlength{\labelwidth}{1em}
     \setlength{\labelsep}{0.5em} } }
\newcommand{\squishend}{
  \end{list}  }

\begin{document}
\title{Endure: A Robust Tuning Paradigm for LSM Trees Under Workload Uncertainty}
 
\author{Andy Huynh}
\affiliation{%
    \institution{Boston University}
}
\email{ndhuynh@bu.edu}

\author{Harshal A. Chaudhari}
\affiliation{%
    \institution{Boston University}
}
\email{harshal@bu.edu}

\author{Evimaria Terzi}
\affiliation{%
    \institution{Boston University}
}
\email{evimaria@bu.edu}

\author{Manos Athanassoulis}
\affiliation{%
    \institution{Boston University}
}
\email{mathan@bu.edu}


\begin{abstract}
Log-Structured Merge trees (LSM trees) are increasingly used as the storage 
    engines behind several data systems, frequently deployed in the 
    cloud. 
Similar to other database architectures, LSM trees take into account
    information about the \emph{expected} workload (e.g., reads vs. writes,
    point vs. range queries) to optimize their performance via     tuning.
Operating in shared infrastructure like the cloud, however, comes with a degree of workload \emph{uncertainty} due 
    to multi-tenancy and the fast-evolving nature of modern applications.
Systems with static tuning discount the variability of such hybrid workloads
    and hence provide an inconsistent and overall suboptimal performance.

To address this problem, we introduce {\Endure} -- a new paradigm for tuning LSM 
    trees in the presence of workload uncertainty.
Specifically, we focus on the impact of the choice of compaction policies, 
    size-ratio, and memory allocation on the overall performance.
{\Endure} considers a robust formulation of the throughput maximization problem,
    and recommends a tuning that maximizes the worst-case throughput over a 
    \emph{neighborhood} of each  expected workload.
Additionally, an uncertainty tuning parameter controls the size of this 
    neighborhood, thereby allowing the output tunings to be conservative or
    optimistic.
Through both model-based and extensive experimental evaluation of {\Endure} in
    the state-of-the-art LSM-based storage engine, RocksDB, we show that the
    robust tuning methodology consistently outperforms classical tuning
    strategies. 
We benchmark {\Endure} using 15 workload templates that generate more than 10000
    unique \emph{noisy} workloads.
The robust tunings output by {\Endure} lead up to a 5{$\times$} improvement in throughput
    in presence of uncertainty. 
On the flip side, when the observed workload exactly matches the
    expected one, {\Endure} tunings have negligible performance loss.

\end{abstract}

\maketitle

\pagestyle{\vldbpagestyle}
\begingroup
\renewcommand\thefootnote{}\footnote{\noindent
This work is licensed under the Creative Commons BY-NC-ND 4.0 International License. Visit \url{https://creativecommons.org/licenses/by-nc-nd/4.0/} to view a copy of this license. For any use beyond those covered by this license, obtain permission by emailing \href{mailto:info@vldb.org}{info@vldb.org}. Copyright is held by the owner/author(s). Publication rights licensed to the VLDB Endowment. \\
\raggedright Proceedings of the VLDB Endowment, Vol. \vldbvolume, No. \vldbissue\ %
ISSN 2150-8097. \\
\href{https://doi.org/\vldbdoi}{doi:\vldbdoi} \\
}\addtocounter{footnote}{-1}\endgroup

\ifdefempty{\vldbavailabilityurl}{}{
\vspace{.3cm}
\begingroup\small\noindent\raggedright\textbf{PVLDB Artifact Availability:}\\
The source code, data, and/or other artifacts have been made available at \url{\vldbavailabilityurl}.
\endgroup
}

\section{Introduction}
\label{sec:introduction}

\begin{figure}[t]
    \centering
    \includegraphics[scale=0.5]{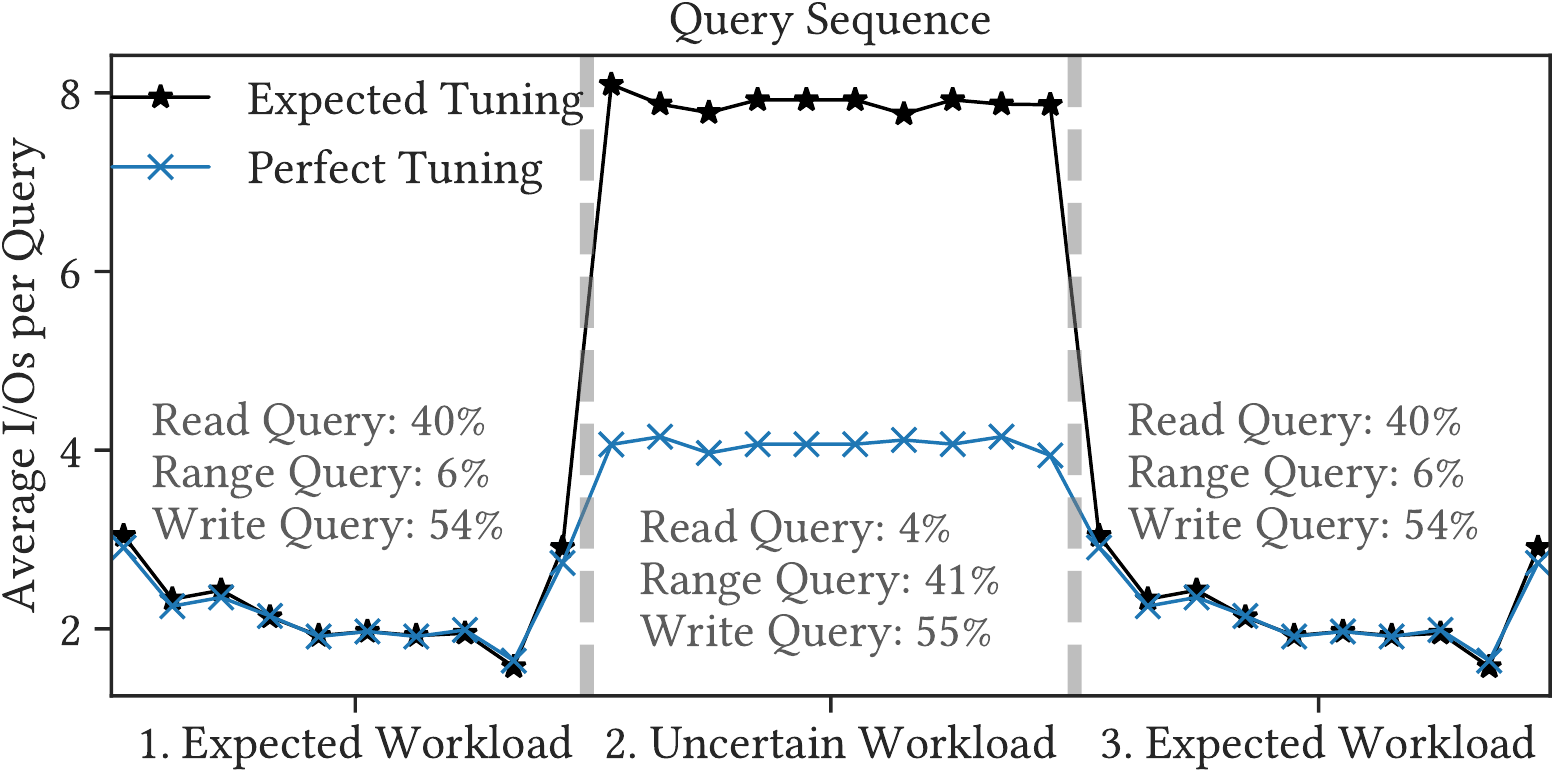}
    \caption{LSM tree tunings and their performance on observed workloads. While
        both workloads have a similar ratio of reads and writes, the uncertain
        workload has a higher percentage of range queries leading to the
        expected system tuning to experience a $2\times$ degradation in
        performance.}
    \label{fig:introduction_query_seq}
\end{figure}

\Paragraph{Ubiquitous LSM-based Key-Value Stores}
Log-Structured Merge trees (LSM trees) is the most commonly deployed data 
    structure used in the backend storage of modern key-value stores
    ~\cite{ONeil1996}.
LSM trees offer high ingestion rate and fast reads,
    making them widely adopted by systems such as RocksDB~\cite{FacebookRocksDB} 
    at Facebook, LevelDB~\cite{GoogleLevelDB} and BigTable~\cite{Chang2006} at 
    Google, HBase~\cite{HBase2013}, Cassandra~\cite{ApacheCassandra} at 
    Apache, WiredTiger~\cite{WiredTiger} at MongoDB, X-Engine \cite{Huang2019} 
    at Alibaba, and DynamoDB~\cite{DeCandia2007} at Amazon.
           
LSM trees store incoming data in a memory buffer, which 
    is flushed to storage when it is full and merged with
    earlier flushed buffers to form a collection of sorted runs 
    with exponentially increasing sizes~\cite{Luo2020b}. 
Frequent merging of sorted runs leads to higher merging costs but facilitates
    faster lookups (\emph{leveling}).
On the flip side, lazy merging policies (\emph{tiering}) trade lookup 
    performance for lower merging costs~\cite{Sarkar2021c}.
Maintaining a separate Bloom filter~\cite{Bloom1970} per sorted run 
    optimizes point queries by avoiding unnecessary accesses to runs 
    that do not contain the desired data.
Range query performance does not benefit from the presence of 
    Bloom filters, yet, it depends on the LSM tree structure.  
   
\Paragraph{Tuning LSM trees} 
As the number of applications relying on 
    LSM-based storage backends increases, the problem 
    of performance tuning for LSM trees has garnered a lot of attention.
A common assumption made by all these methods is that one has \emph{complete knowledge about the 
    expected workload and the execution environment}.
Given such knowledge, prior work optimizes for the memory allocation 
    to Bloom filters across different levels, memory distribution between 
    the buffers and the Bloom filters, and the choice of merging policies 
    (i.e., \emph{leveling} or \emph{tiering})~\cite{Dayan2017,Dayan2018a}.  
Different optimization objectives have led to hybrid merging policies
    with more fine-grained tunings~\cite{Dayan2018,Dayan2019,Idreos2019}, 
    optimized memory allocation strategies~\cite{Bortnikov2018,Kim2020,Luo2020a}, 
    variations of Bloom filters~\cite{Luo2020,Zhang2018a,Zhang2020a}, new 
    compaction routines~\cite{Alkowaileet2020,Luo2019a,Sarkar2020,Sarkar2021c,Zhang2020}, 
    and exploitation of data characteristics
    ~\cite{Yang2020, Ren2017, Absalyamov2018}.

\Paragraph{The Only Certainty is Uncertainty}
Even when accurate information about the workload and underlying hardware is 
    available, tuning data systems is a notoriously difficult 
    research problem \cite{Chaudhuri2004a,Chaudhuri2005,Shasha2002}. 
The explosive growth in the public and private use of the cloud infrastructure 
    for data management~\cite{Hayes2008,Intel2011,GrandViewResearch2019} 
    has exacerbated this problem because of increased uncertainty and
    variability in workloads
    ~\cite{Chohan2010,Galante2012,Herbst2013,Holze2010,Mohan2016,Ozcan2017,
    Pezzini2014,Schnaitter2006,Schnaitter2007, Schnaitter2012,Wolski2017}.

\Paragraph{An Example}  
Before describing our framework, we give an example that depicts how
    variation in observed workloads relative to the expected workload -- 
    used during tuning of the LSM tree-based storage -- leads to suboptimal 
    performance. 
In Figure~\ref{fig:introduction_query_seq}, the $x$-axis shows a sequence of 
    workloads executed over an LSM-based engine, while the $y$-axis shows the 
    average disk accesses per workload. 
The experiment is split into three sessions -- the first and the last sessions 
    receive the expected workload,
    while the second session receives a different workload.
Although it has the same reads vs. writes ratio as the expected 
    workload, it has a higher percentage of short range queries in comparison
    to the point queries.
The solid black line shows the performance of a system tuned for the expected 
    workload. 
Note that the average I/Os increase dramatically in the second session 
    even though the amount of data being read is approximately the same.
On the other hand, the blue line corresponds to each session having its ideal 
    tuning, leading to only half as many I/Os per operation.
Note that it is not feasible to continually change tunings during execution, 
    as it requires redistribution of the allocated memory between different 
    components of the tree and potentially changing its shape. 
Hence, we want to \emph{find a tuning that is close-to-optimal for both the 
    expected and the observed workload}.
    
\Paragraph{Our Work: Robust LSM Tree Tuning}
To address this suboptimality caused by variations in the 
    observed workload, we depart from the classical view of database tunings 
    which assumes accurate knowledge about the expected workload. 
Rather, we introduce {\Endure}, a new \emph{robust tuning paradigm} that
    incorporates expected uncertainty into optimization and apply it to LSM
    trees.

We propose a problem formulation that seeks an LSM tree configuration that
    maximizes the worst-case throughput over all the workloads in the
    \emph{neighborhood} of an expected workload and call it the {\robustw}
    problem. 
We use the notion of KL-divergence between probability distributions to define
    the neighborhood size, implicitly assuming that the uncertain
    workloads would be contained in the neighborhood. 
As the KL-divergence boundary condition approaches zero, our problem becomes 
    equivalent to the classical optimization problem (referred henceforth 
    as the {\nominal} problem). 
More specifically, our approach uses as input the expected size of the 
    uncertainty neighborhood, which dictates the qualitative 
    characteristics of the solution. 
Intuitively, the larger the size of the uncertainty region, the larger the
    workload discrepancy a robust tuning can absorb.
Leveraging work on robust optimization from the Operations Research 
    community~\cite{Bertsimas2010, Ben-Tal1998, Ben-Tal2013}, we efficiently
    solve the {\robustw} problem and find the robust tuning for LSM tree-based
    storage systems.
A similar problem of using workload uncertainty while determining the physical design
    of column-stores has been explored in prior work~\cite{Mozafari2015}.
However, their methodology is not well suited for the LSM tuning problem. 
We provide additional details regarding this in Section~\ref{sec:related_work}.

\Paragraph{Contributions}
To the best of our knowledge, our work presents the first systematic approach
    for robust tuning of LSM tree-based key-value stores under workload 
    uncertainty. 
Our technical and practical contributions can be summarized as follows: 
\begin{itemize}[leftmargin=1.3em, itemsep=0.22em]
    \item We incorporate workload uncertainty in LSM tuning and show
        how to find a robust tuning efficiently. Our algorithm can be tuned
        for varying uncertainty and is simple enough to be adopted by current
        state-of-the-art LSM storage engines (\S\ref{sec:algo-for-robust-tuning}
        and \S\ref{sec:tuning_lsm_trees}).
    \item We augment existing analytical cost models of LSM tree-based 
        storage engines with more precise estimates of workload execution costs
        (\S\ref{sec:tuning_lsm_trees}).
    \item In our model-based analysis, we show that robust tunings from
        {\Endure} provide up to 5$\times$ higher throughput when faced with
        uncertain workloads, and match classical tuning performance when there
        is no uncertainty (\S\ref{sec:model-evaluation}).
    \item We integrate {\Endure} into RocksDB, a state-of-the-art LSM storage
        engine, where we show both system I/O and latency reductions of up to
        $90\%$. Additionally, we show that {\Endure} scales with database size
        (\S\ref{sec:system-evaluation}).
    \item To encourage reproducible research, we make our robust tuning framework 
        publicly available~\cite{github-repo}.
\end{itemize}

\begin{figure}[t]
    \centering
    \includegraphics[scale=0.75]{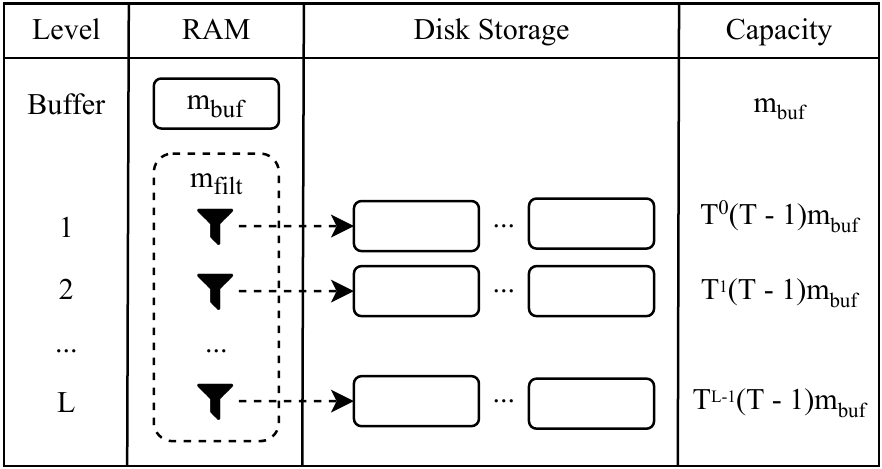}
    \caption{Overview of the structure of an LSM tree}
    \label{fig:lsm-overview}
\end{figure}

\section{Background on LSM Trees}
\label{sec:background}

\Paragraph{Basics} 
LSM trees use the \textit{out-of-place} ingestion paradigm to store key-value
    pairs. 
Writes, updates, or deletes are buffered in a memory buffer, and once
    full, its contents are sorted based on the key, forming an
    \textit{immutable sorted run}. 
This run is then flushed to the first level on secondary storage. 
Each level of sorted runs has a tunable maximum permitted size. 
Overall, for an LSM tree with $L$ disk-resident levels, we denote the
    memory buffer as Level $0$, and the remaining
    levels in storage $1$ to $L$. 
The disk-resident sorted runs have exponentially increasing sizes following a 
    tunable size ratio $\sizeratio$.
Figure~\ref{fig:lsm-overview} shows an overview of an LSM tree.

We denote the number of bits of main memory allocated to the buffer as $\mbuf$,
    which holds a number of entries with fixed entry size $E$. 
For example, in RocksDB the default buffer size is $\mbuf=64$MB, and depending 
    on the application, the entry size typically varies between 64B and 1KB. 

Level 0 can be updated in-place as it is in memory, however, runs in
    levels 1 and beyond are immutable. 
Each level $i$ has a capacity threshold of 
    $(\sizeratio - 1) T^{i-1} \cdot \frac{\mbuf}{E}$ entries, thus, level
    capacities are exponentially increasing by a factor of $\sizeratio$.  
The total number of levels $L$ for a given {\sizeratio} is
\begin{equation} 
L(T) = \Bigg\lceil \log_T \left( {\frac{N \cdot E}{\mbuf}} + 1 \right) \Bigg\rceil ,
\label{eq:levels}
\end{equation}
where $N$ is the total number of entries across all levels~\cite{Dayan2018a, Luo2020, Sarkar2020}.

\Paragraph{Compaction Policies: Leveling and Tiering} 
Classically, LSM trees support two merging policies: leveling and tiering. 
In leveling, each level may have at most one run, and every time a run in Level
    $i - 1$ ($i \geq 1$) is moved to Level $i$, it is greedily sort-merged
    (compaction) with the run from Level $i$, if it exists. 
With tiering, every level must accumulate $\sizeratio$ runs before they trigger
    a compaction. 
During a compaction, entries with a matching key are consolidated and only the
    most recent valid entry is retained~\cite{Dong2017, ONeil1996}. 
Recently hybrid compaction policies fuse leveling and tiering in a single tree
    to strike a balance between the read and write throughput~\cite{Dayan2018,
    Dayan2019}.

\Paragraph{LSM tree Operations}
An LSM tree supports: (a) writes of new key-value pairs, (b) point queries, and
    (c) range queries.

\emph{Writes:} 
A write operation is handled by a buffer append, and if the buffer gets full, 
    it triggers a compaction. 
Any write may include either a new key-value pair, an existing key that \emph{updates} its value, or a special entry 
    that \emph{deletes} an existing key.

\emph{Point Queries:} 
A point query searches for the value of a specific unique key. 
It begins by looking at the memory buffer, then traverses the tree from the 
    smallest to the largest level. For tiering, within a level, a lookup 
    moves from the most to the least recent tier. 
The lookup terminates when it finds the first matching entry. 
Note that a point query might return an \emph{empty} or a
    \emph{non-empty} result. 
We differentiate the two because, in general, workloads with empty point 
    queries can be further optimized \cite{Dayan2017,Dayan2018a}.

\emph{Range Queries:}
A range lookup returns the most recent versions of the target keys 
    by sort-merging all qualifying runs from the tree.

\Paragraph{Optimizing Lookups} 
Read performance is optimized using Bloom filters and fence pointers. 
In the worst case, a lookup needs to probe every run. 
To reduce this cost, LSM engines use one Bloom filter per run in main memory
    ~\cite{Dayan2017, FacebookRocksDB}. 
Bloom filters \cite{Bloom1970} are probabilistic membership test data structures
    that exhibit a false positive $f$ as a function of the ratio between the
    memory allocated $\mfilt$ to them and the elements it indexes.
In LSM trees, Bloom filters allow a lookup to skip probing a run altogether if
    the filter-lookup returns negative.
In practice, for efficient storage, Bloom filters are maintained at the
    granularity of files~\cite{Dong2017}. 
Fence pointers store the smallest key per disk page in memory~\cite{Dayan2017},
    to quickly identify which page(s) to read for a lookup, and perform up to
    one I/O per run for point lookups.

\Paragraph{Tuning LSM Trees} 
Prior to this work, efforts to systematically tune LSM trees assume that the
    workload information and the execution environment are accurately known.
Under that assumption, the main focus on LSM tuning has been on deciding how to
    allocate the available main memory between Bloom filters and buffering
    ~\cite{Dayan2017,Kim2020,Luo2020a}, while often the size ratio and the 
    merging strategy was also co-tuned~\cite{Dayan2018a}.
Such design decisions are common across industry standard LSM-based engines
    such as Apache Cassandra~\cite{ApacheCassandra},
    AsterixDB~\cite{Alsubaiee2014}, RocksDB~\cite{RocksDB2020}, and
    InfluxDB~\cite{InfluxDB}.
In addition, recent work has introduced new hybrid merging strategies 
    ~\cite{Dayan2018,Dayan2019,Sarkar2021c}, and optimizations for faster data 
    ingestion~\cite{Luo2019b} and performance stability~\cite{Luo2019a}.

\section{Problem Definitions}
In this section, we provide the formal problem definitions on how to choose the \emph{design parameters} of 
an LSM tree. Before proceeding, we give a brief introduction to our notation.

\subsection{Notation}\label{sec:notation}
As we discussed above, LSM trees have two types of parameters: the \emph{design
    parameters} that are changed primarly for performance, and the \emph{system
    parameters} that are given and therefore untunable.

\Paragraph{Design Parameters}
The design parameters we consider in this paper are the size-ratio
    ({\sizeratio}), the memory allocated to the Bloom filters ({\mfilt}), the
    memory allocated to the write buffer ({\mbuf}) and the compaction policy
    ({\policy}).
These are ubiquitous design parameters and have been extensively studied as
    having the largest impact on performance~\cite{Dayan2017, Dayan2018a,
    Luo2020b}.
Therefore, we focus on these parameters in order to define a problem that is
    agnostic to the LSM engine used.
Recall that the policy refers to either leveling or tiering, as discussed in the
    previous section.

\Paragraph{System Parameters}
A complex data structure like an LSM tree also has various \emph{system
    parameters} and other non-tunable ones (e.g., total memory ({\mtot}), data
    entry size $E$, page size $B$, data size $N$).

\Paragraph{LSM Tree Configuration}
For notation, we use $\configuration$ to denote the LSM tree tuning configuration
    which describes the values of the tunable parameters together
    $\configuration := (\sizeratio, \mfilt, \policy)$.
Note that we only use the memory for Bloom filters $\mfilt$ and not $\mbuf$,
    because the latter can be derived using the former and total available
    memory: $\mbuf=\mtot-\mfilt$.

\Paragraph{Workload}
The choice of the parameters in $\configuration$ depends on the input (expected)
    workload, i.e., the fraction of empty lookups ({\emptylookup}), non-empty
    lookups ({\nonemptylookup}), range lookups ({\range}), and write ({\update})
    queries.
Note that this workload representation is common for analyzing and tuning LSM
    trees~\cite{Dayan2018a, Luo2020b}.
Additionally, complex workloads (i.e., SQL statements) generate access patterns
    on the storage engine and can be broken down into the same basic
    operations.
This mapping of complex queries to basic operations is also common for
    performance tuning of LSM tree-based storage engines~\cite{Cao2020}.
Therefore a workload can be expressed as a vector 
    ${\workload = (\emptylookup, \nonemptylookup, \range, \update)^\intercal \geq 0}$
    describing the proportions of the different kinds of queries.
Clearly, $\emptylookup+\nonemptylookup+\range+\update = 1$ or alternatively:
    $\workload^\intercal \columnvec = 1$ where {\columnvec} denotes a column vector of
    ones. 

\begin{table}[t]\centering
\resizebox{\columnwidth}{!}{
\begin{tabular}{ccl}
    \toprule
    Type & Term & Definition \\
    \toprule
        Design & $\mfilt$ &  Memory allocated for Bloom filters  \\ 
               & $\mbuf$ &  Memory allocated for the write buffer \\ 
               & $\sizeratio$ &  Size ratio between consecutive levels   \\ 
               & \policy & Compaction policy (\emph{tiering}/\emph{leveling}) \\ 
    \midrule
        System & $\mtot$ & Total memory (filters+buffer) ($\mtot=\mbuf+\mfilt$)\\
               & $E$ &  Size of a key-value entry  \\ 
               & $B$ &  Number of entries that fit in a page   \\ 
               & $N$ & Total number of entries    \\ 	
    \midrule
        Workload & $z_0$ & Percentage of zero-result point lookups\\ 
                 & $z_1$ & Percentage of non-zero-result point lookups\\ 
                 & $q$ & Percentage of range queries\\ 
                 & $w$ & Percentage of writes\\ 
    \bottomrule
\end{tabular}
}
\caption{Summary of problem notation}
    \label{tab:params}	
\end{table}

Each type of query (non-empty lookups, empty lookups, range lookups
    and writes) has a different cost, denoted as $Z_0(\configuration)$,
    $Z_1(\configuration)$, $Q(\configuration)$, $W(\configuration)$, as there is
    a dependency between the cost of each type of query and the design
    ${\configuration}$.
For ease of notation, we use 
    $\costvec(\configuration) = \left(Z_0(\configuration), Z_1(\configuration), Q(\configuration), W(\configuration)\right)^\intercal$ 
    to denote the vector of the costs  of executing
    different types of queries.
Thus, given a specific configuration ({\configuration}) and a workload ({\workload}),
    the expected cost for the workload can be computed as:
{%
\begin{equation}
    \label{eq:thecost}
    \cost(\workload, \configuration) = \workload^\intercal
    \costvec(\configuration)=\nonemptylookup \cdot
    Z_0(\configuration)+\emptylookup \cdot Z_1(\configuration) + \range\cdot Q(\configuration) + \update \cdot W(\configuration).
\end{equation}
}%

\noindent We present a summary of all of our notation in Table~\ref{tab:params}.

\subsection{The Nominal Tuning Problem}
Traditionally, the designers have focused on finding the 
configuration ${\configuration}^\ast$ that minimizes the total cost 
$\cost(\workload, \configuration^\ast)$,
 for a given fixed workload $\workload$.  We call this problem the
 {\nominal} problem, defined as follows:
 
 \begin{problem}[{\nominal}]\label{problem:nominal}
 Given fixed $\workload$  find the tuning configuration of the LSM tree $\configuration_N$ such that
\begin{equation}
\label{eq:nominal_problem}
    \configuration_N = \argmin_{\configuration} \cost(\workload, \configuration).
\end{equation}
\end{problem}

\noindent The nominal tuning problem described above captures the classical
    tuning paradigm. It uses a cost-model to find a system configuration that 
    minimizes the cost given a specific workload and system environment.
    Specifically, prior tuning approaches for LSM trees solve the nominal tuning
    problem when proposing optimal memory allocation, and merging policies
    \cite{Dayan2017,Dayan2018a,Luo2020a}.

\subsection{The Robust Tuning Problem}
\label{subsec:robust-tuning}

In this work, we attempt to compute high-performance configurations that minimize
    the expected cost of operation, as expressed in Equation~\eqref{eq:thecost}, 
    in the presence of uncertainty with respect to the expected workload.

The {\nominal} problem assumes perfect information about the 
    workload for which to tune the system. For example, we may assume that the
    input vector $\workload$ represents the workload for which we 
    optimize, while in practice, $\workload$ is simply an estimate of what the
    workload will look like.
Hence, the configuration obtained by solving Problem~\ref{problem:nominal} may
    result in high variability in the system performance that will inevitably
    depend on the actual observed workload upon the deployment of the system. 
    
We capture this uncertainty by reformulating Problem~\ref{problem:nominal} to
    take into account the variability that can be observed in the input workload. 
Given an expected workload $\workload$, we introduce the notion of the
    \emph{uncertainty region} of $\workload$, which we denote by
    $\mathcal{U}_\workload$.

We can define the robust version of Problem~\ref{problem:nominal}, under the 
    assumption that there is uncertainty in the input workload as follows:

\begin{problem}[{\robustw}]\label{problem:robustw} 
Given $\workload$ and uncertainty region $\mathcal{U}_\workload$ 
find the tuning configuration of the LSM tree $\configuration_R$ such that
\begin{eqnarray}
\label{eq:robust_workload_problem}
\configuration_R &=& \argmin_{\configuration} \cost(\obsworkload,
    \configuration) \nonumber\\
    \textrm{s.t.,}&& \obsworkload \in \mathcal{U}_{\workload}.
\end{eqnarray}
\end{problem}

\noindent Note that the above problem definition intuitively states the following: it
    recognizes that the input workload $\workload$ won't be observed exactly,
    and it assumes that any workload in $\mathcal{U}_\workload$ is possible.
    Then, it searches for the configuration $\configuration_\workload$ that is
    best for the \emph{worst-case} scenario among all those in
    $\mathcal{U}_\workload$. 

The challenge in solving {\robustw}  is that one needs to explore all the
    workloads in the uncertainty region in order to solve the problem.  In the
    next section, we show that this is not necessary.  In fact, by appropriately
    rewriting the problem definition we show that we can solve
    Problem~\ref{problem:robustw} in polynomial time.

\section{Algorithms for ROBUST TUNING}
\label{sec:algo-for-robust-tuning}

In this section, we discuss our solutions to the {\robustw} problem. 
On a high level, the solution strategy is the following:
    first, we express the objective of the problem (as expressed in 
    Equation~\eqref{eq:robust_workload_problem}) as a standard continuous
    optimization problem.  
We then take the \emph{dual} of this problem and use existing results in robust 
    optimization to show: $(i)$ the duality gap between the primal and the dual
    is zero, and $(ii)$ the dual problem is solvable in polynomial time.
Thus, the dual solution can be translated into the optimal solution for the 
    primal, i.e., the original {\robustw} problem.  
The specifics of the methodology are described below: 

\Paragraph{Defining the Uncertainty Region $\mathcal{U}_\workload$}  
Recall that $\workload$ is a probability vector, i.e., 
    $\workload^\intercal\columnvec = 1$. Thus, in order to define the 
    uncertainty region $\mathcal{U}_\workload$, we use the Kullback-Leibler (KL) 
    divergence function~\cite{Kullback1951}. KL-divergence for two probability
    distributions is defined as follows:
 
\begin{definition}
\label{def:KL}   
    The KL-divergence distance between two vectors
    $\vec{p} = (p_1, \cdots, p_m)^\intercal \geq 0$ and
    $\vec{q} = (q_1, \cdots, q_m)^\intercal \geq 0$ in 
    $\mathbb{R}^m$ is defined as,
\begin{equation*}
    I_{KL}(\vec{p}, \vec{q}) = \sum_{i=1}^{m}p_i \log\bigg(\frac{p_i}{q_i}\bigg).
\end{equation*}
\end{definition}
\noindent Note that we could have potentially used other divergence 
    functions~\cite{Pardo2018-ou} instead of the KL-divergence.
We use the KL-divergence as we believe it fits our goal and intuitive
    understanding of the space of workloads.

Using the notion of KL-divergence we can now formalize the uncertainty
    region around an expected workload {\workload} as follows,
    \begin{equation}\label{eq:workloaduncertaintyregion}
        \mathcal{U}_{\workload}^\rho = \{\obsworkload \in \mathbb{R}^4\textrm{
            }|\textrm{  } \obsworkload
    \geq 0, \obsworkload^\intercal\columnvec = 1, I_{KL}(\obsworkload,
    \workload) \leq \rho\}.
    \end{equation}
   
\noindent Here, $\rho$ determines the maximum KL-divergence that is allowed between any
    workload $\obsworkload$ in the uncertainty region and the input expected
    workload $\workload$. 
Note that the definition of the uncertainty region takes as input the parameter
    $\rho$, which intuitively defines the neighborhood around the expected
    workload. 
This $\rho$ can be computed as the mean KL-divergence from the historical
    workloads.  

In terms of notation, $\rho$ input is required for defining the uncertainty
    region $\mathcal{U}_\workload^\rho$.
However, we drop the superscript notation unless required for context.

\Paragraph{Rewriting of the ROBUST TUNING Problem (Primal)}
Using the above definition of the workload uncertainty region 
    $\mathcal{U}_\workload^\rho$, we are now ready to proceed to the solution of
    the {\robustw} problem. 
For a given $\rho$, the problem definition as captured by
    Equation~\eqref{eq:robust_workload_problem} can be rewritten as follows:
    \begin{equation}\label{eq:robust_tuning_problem1}
        \min_{\configuration} \max_{\obsworkload \in \mathcal{U}_{\workload}^\rho}
        \obsworkload^\intercal \costvec(\configuration).
    \end{equation}

\noindent Note that the above equation is a simple rewrite of
    Equation~\eqref{eq:robust_tuning_problem1} that captures the intuition that
    the optimization is done over the \emph{worst-case} workload among all the
    workloads in the uncertainty region $\mathcal{U}_\workload$.
An equivalent way of writing Equation~\eqref{eq:robust_tuning_problem1} is by
    introducing an additional variable $\beta\in\mathbb{R}$, then writing the
    following:
\begin{eqnarray}
\label{eq:robust_counterpart_workload_problem}
\min_{\beta, \configuration}&\beta& \nonumber\\
    \textrm{s.t.,}&\obsworkload^\intercal\costvec(\configuration) \leq \beta
                  &\forall  \obsworkload \in \mathcal{U}_{\workload}.
\end{eqnarray}
This reformulation allows us to remove the $\min\max$ term in the objective from 
    Equation~\eqref{eq:robust_tuning_problem1}. 
The constraint in Equation~\eqref{eq:robust_counterpart_workload_problem}
    can be equivalently expressed as,
\begin{eqnarray*}
\label{eq:inner_optimization}
    \beta &\geq&
    \max_{\obsworkload}\big\{\obsworkload^\intercal\costvec(\configuration) | \obsworkload \in
    \mathcal{U}_{\workload}\big\}\nonumber\\
    &=&\max_{\obsworkload \geq
0}\bigg\{\obsworkload^\intercal\costvec(\configuration)\bigg|\obsworkload^\intercal\columnvec
= 1, \sum_{i=1}^{m} \hat{w}_i \log\bigg(\frac{\hat{w}_i}{w_i}\bigg) \leq \rho\bigg\}.
\end{eqnarray*}
Finally, the Lagrange function for the optimization on the right-hand side of 
    the above equation is:
\begin{equation*}
    \mathcal{L}(\obsworkload, \lambda, \eta) =
    \obsworkload^\intercal\costvec(\configuration) + \rho \lambda - \lambda
    \sum_{i=1}^{m}\hat{w}_i \log\bigg(\frac{\hat{w}_i}{w_i}\bigg) + \eta(1 -
    \obsworkload^\intercal \columnvec),
\end{equation*}
where $\lambda$ and $\eta$ are the Lagrangian variables. 

\Paragraph{Formulating the Dual Problem}
We can now express the dual objective as,
\begin{equation}\label{eq:dual}
    g(\lambda, \eta) = \max_{\obsworkload \geq 0}\mathcal{L}(\obsworkload,
    \lambda, \eta),
\end{equation}
which we need to \emph{minimize}.

Now we borrow the following result from ~\cite{Ben-Tal2013-kw},
\begin{lemma}[\cite{Ben-Tal2013-kw}]
A configuration {\configuration} is the optimal solution to the {\robustw} 
    problem if and only if $\min_{\eta, \lambda \geq
    0}g(\lambda, \eta) \leq \beta$ where the minimum is attained for some value 
    of $\lambda \geq 0$.
\end{lemma}
In other words, minimizing the dual objective $g(\lambda, \eta)$ (as expressed
    in Equation~\eqref{eq:dual}) will lead to the optimal solution for the
    {\robustw} problem. 

\Paragraph{Solving the Dual Optimization Problem Optimally} 
Formulating the dual problem and using the results of 
    Ben-Tal~\etal~\cite{Ben-Tal2013-kw}, we have shown that the dual solution 
    leads to the optimal solution for the {\robustw} problem.
Moreover, we can obtain the optimal solution
    to the original {\robustw} problem in polynomial time, a consequence of the
    tractability of the dual objective. 

To solve the dual problem, we first simplify the dual objective 
    $g(\lambda,\eta)$ so that it takes the following form:
\begin{equation}\label{eq:final_dual_rewrite}
 g(\lambda,\eta) = \eta + \rho \lambda + \lambda
    \sum_{i=1}^{k}w_i\phi_{KL}^* \bigg(\frac{\costvec_i(\configuration) - \eta}{\lambda}\bigg).
\end{equation}

\noindent In Equation~\eqref{eq:final_dual_rewrite}, $\phi_{KL}^*(.)$ denotes the 
    conjugate of KL-divergence function and $\costvec_i$ corresponds to the 
    $i$-th dimension of the cost vector $\costvec(\configuration)$ as defined 
    in Section~\ref{sec:notation} -- clearly in this case $k=4$ as we have $4$ 
    types of queries in our workload.
Results of Ben-Tal~\etal~\cite{Ben-Tal2013-kw} show that minimizing the dual 
    function as described in Equation~\eqref{eq:final_dual_rewrite} is a convex 
    optimization problem, and it can be solved optimally in polynomial time if 
    and only if the cost function $\costvec(\configuration)$ is convex in all 
    its dimensions.

In our case, the cost function for the range queries is not convex w.r.t.
    size-ratio {\sizeratio} for the tiering policy.
However, on account of its smooth non-decreasing form, we are still able
    to find the global minimum solution for 
\begin{eqnarray}
\min_{\configuration, \lambda \geq 0, \eta}
\bigg\{\eta + \rho \lambda + \lambda
    \sum_{i=1}^{m}w_i\phi_{KL}^* \bigg(\frac{c_i(\configuration) -
\eta}{\lambda}\bigg)\bigg\}.
\end{eqnarray}
This minimization problem can be solved using the 
    Sequential Least Squares Quadratic Programming solver (SLSQP) included in 
    the popular Python optimization library SciPy \cite{2020SciPy-NMeth}.
Solving this problem outputs the values of the Lagrangian variables $\lambda$ 
    and $\eta$ and most importantly the configuration $\configuration$ that 
    optimizes the objective of the {\robustw} problem -- for input $\rho$.
In terms of running time, SLSQP solver outputs a robust tuning configuration 
    for a given input in less than a second.

\section{The Cost Model of  LSM Trees}
\label{sec:tuning_lsm_trees}

In this section, we provide the detailed cost model used in {\Endure} to
    accurately capture the behavior of an LSM tree. 
Following prior work on LSM trees \cite{Dayan2018a, Luo2020}, we focus on four 
    types of operations: point queries that return an empty result, point 
    queries that have a match, range queries, and writes.

\subsection{Model Basics}

When modeling the read cost of LSM trees, a key quantity to accurately capture 
    is the amount of extra read I/Os that take place. 
While Bloom filters are used to minimize those, they allow a small fraction of 
    false positives. 
In particular, a point lookup probes a run's filter before accessing the run in 
    secondary storage. 
If the filter returns negative, the target key does not exist in the run, and 
    so the lookup skips accessing the run and saves one I/O.
If a filter returns positive, then the target key may exist in the run, so the 
    lookup probes the run at a cost of one I/O.
If the run actually contains the key, the lookup terminates.
Otherwise, we have a \emph{false positive} and the lookup continues to probe 
    the next run.
False positives increase the I/O cost of lookups. 
The false positive rate ($\epsilon$) of a standard Bloom filter designed to 
query $\mathbf{n}$ entries using a bit-array of size $\mathbf{m}$ is shown by 
    \cite{Tarkoma2012} to be calculated as follows:
\begin{equation*} 
\label{eq:bloom}
    \epsilon = e^{-\frac{\mathbf{m}}{\mathbf{n}} \cdot \ln(2)^2}.
\end{equation*}
Note that the above equation assumes the use of an optimal number of hash 
    functions in the Bloom filter~\cite{enwiki:1025193696}.

Classically, LSM tree based key-value stores use the same number of 
    bits-per-entry across all Bloom filters. This means that a lookup probes 
    on average $O\left(e^{-\nicefrac{\mfilt}{N}}\right)$ of the runs, 
    where $\mfilt$ is the overall amount of main memory allocated to the 
    filters. 
As $\mfilt$ approaches 0 or infinity, the term 
    $O\left(e^{-\nicefrac{\mfilt}{N}}\right)$
    approaches 1 or 0 respectively.  
Here, we build on of the state-of-the-art Bloom filter allocation 
    strategy proposed in Monkey \cite{Dayan2017,Dayan2018a} that uses different 
    false positive rates at different levels of the LSM tree to offer  
    optimal memory allocation; for a size ratio {\sizeratio}, the false positive 
    rate corresponding to the Bloom filter at the level $i$ is given by
\begin{equation}
\label{eq:bloom2}
    f_i(T) = \frac{T^{\frac{T}{T-1}}}{T^{L(T)+1-i}}\cdot e^{-\frac{\mfilt}{N}\ln(2)^2}.
\end{equation}

\noindent Additionally, false positive rates for all levels satisfy $0 \leq f_i(T) \leq 1$.
It should be further noted that Monkey optimizes false positive rates at
    individual levels to minimize the worst case average cost of empty point
    queries.
Non-empty point query costs, being significantly lower than those of empty point
    queries, are not considered during the optimization process.

\Paragraph{LSM Tree Design \& System Parameters}
In Section \ref{sec:notation} we introduced the key design and system parameters
    needed to model LSM tree performance. 
In addition to those parameters, there are two auxiliary and derived parameters
    we use in the cost model presented in this section: the potential
    storage asymmetry in reads and writes (\asym) and the expected selectivity
    of range queries (\querySel).

\subsection{The Cost Model}
In this section, we model the costs in terms of expected number of I/O 
    operations required for the fulfillment of the individual queries.

\Paragraph{Expected Empty Point Query Cost ($Z_0$)} 
A point query that returns an empty result will have to visit all levels (and
    every sorted run of every level for tiering) where false positives in the 
    Bloom filters trigger I/O operations. 
Thus, the expected number of I/O operations per level depends on the Bloom filter 
    memory allocation at that level.  
Hence, Equation \eqref{eq:read} expresses $Z_0$ in terms of the false positive 
    rates at each level:
\begin{equation} 
\label{eq:read}
Z_0(\configuration) = 
\begin{cases}
{\color{white} (T-1)} \sum_{i=1}^{L(T)} f_i(T),	&  \text{if }\policy=\text{leveling} \\[4pt]
(T-1) \sum_{i=1}^{L(T)} f_i(T) ,	&  \text{if }\policy=\text{tiering}.  \\
\end{cases} \\
\end{equation}
In the leveling policy, each level has exactly one run.  
On the other hand, with tiering policy, each level has up to $(T-1)$ runs.
All runs at the same level in tiering have equal false positives rates 
    on account of their equal sizes. 

\Paragraph{Expected Non-empty Point Query Cost ($Z$)}
There are two components to the expected non-empty point query
    cost.
First, we assume that the probability of a point query finding a non-empty 
    result in a level is proportional to the size of the level. 
Thus, the probability of such a query being satisfied on level $i$ by a unit
    cost I/O operation is simply 
    $\frac{(T-1)\cdot T^{i-1}}{N_f(T)}\cdot \frac{\mbuf}{E}$, where $N_f(T)$ denotes the 
    number of entries in a tree completely full upto $L(T)$ levels. Thus,

\begin{equation}
N_f(T) = \sum_{i=1}^{L(T)} (T-1)\cdot T^{i-1}\cdot \frac{\mbuf}{E}. 
\end{equation}

\noindent Second, we assume that all levels preceding level $i$ trigger I/O 
    operations with probability equivalent to the false positive rates of the 
    Bloom filters at those levels.
Similar to the empty point queries, the expected cost of such failed
    I/Os on preceding levels is simply $\sum_{j=1}^{i-1}f_j(T)$.
In the case of tiering, we assume that on average, the entry is found in the
    middle run of the level resulting in an additional
    $\frac{(T-2)}{2}\cdot f_i(T)$ extra I/O operations. 
Thus, we can compute the non-empty point query cost as an expectation over the 
    entry being found on any of the $L(T)$ levels of the tree as follows:

\ssmall
\begin{equation}
\label{eq:read-non-empty}
Z_1(\configuration) =
\begin{cases}
    \sum_{i=1}^{L(T)} \frac{(T-1)\cdot T^{i-1}}{N_f(T)}\cdot \frac{\mbuf}{E} \bigg(1 +
    \sum_{j=1}^{i-1}f_j(T)\bigg), &\text{if } \policy=\text{leveling} \\[8pt]
    \sum_{i=1}^{L(T)} \frac{(T-1)\cdot T^{i-1}}{N_f(T)}\cdot \frac{\mbuf}{E} \bigg(1 +
        (T-1)\cdot\sum_{j=1}^{i-1}f_j(T)\\
    {\color{white}\sum_{i=1}^{L(T)} \frac{(T-1)\cdot T^{i-1}}{N_f(T)}} +
\frac{(T-2)}{2}\cdot f_i(T)\bigg), &\text{if } \policy=\text{tiering}.  \\
\end{cases}
\end{equation}
\normalsize

\Paragraph{Range Queries Cost ($Q$)}
A range query issues $L(T)$ or $L(T) \cdot (\sizeratio-1)$ disk seeks 
    (one per run) for leveling and tiering respectively.
Each seek is followed by a sequential scan. The cumulative number of pages
    scanned over all runs is $\querySel \cdot \frac{N}{B}$, where {\querySel} 
    is the average proportion of all entries included in range lookups.  
Hence, the overall range lookup cost $Q$ in terms of pages reads is as follows: 

\begin{equation} 
\label{eq:range}
Q(\configuration) = 
\begin{cases}
    \querySel \cdot \frac{N}{B} + L(T),  ~~ & \text{if } \policy=\text{leveling}  \\[5pt]
    \querySel \cdot \frac{N}{B} + L(T) \cdot (\sizeratio-1)  , \qquad     & \text{if } \policy=\text{tiering.}  \\
\end{cases}  \\
\end{equation}

\Paragraph{Write Cost ($W$)}
We model worst-case writing cost assuming that the vast majority 
    of incoming entries do not overlap. 
This means that most entries will have to propagate through all levels of the
    LSM tree.
Following the state-of-the-art write cost model, we assume that every 
    written item participated in $\approx \frac{\sizeratio-1}{\sizeratio}$ and
    $\approx \frac{\sizeratio-1}{2}$ merges with tiering and leveling 
    respectively.
We multiply these costs by $L(T)$ since each entry gets merged across all 
    levels, and we divide by the page size $B$ to get the units in terms of 
    I/Os.  
Since LSM trees often employ solid-state storage that has an asymmetric cost for
    reads and writes, we represent this storage asymmetry as $\asym$. 
For example, a device for which a write operation is twice as expensive as a 
    read operation has $\asym=2$.  
The overall I/O cost is captured by Equation \eqref{eq:write}:
\begin{equation} 
\label{eq:write}
W(\configuration) = 
\begin{cases}
    \frac{L(T)}{B} \cdot  \frac{ (\sizeratio-1)}{2  } \cdot (1 + \asym) ,  ~~ & \text{if } \policy=\text{leveling}\\[5pt]
    \frac{L(T)}{B} \cdot  \frac{ (\sizeratio-1)}{\sizeratio  } \cdot (1 + \asym) ,  \qquad    & \text{if }
    \policy=\text{tiering.}  \\
\end{cases}  \\
\end{equation}
When $\sizeratio$ is set to 2, tiering and leveling behave identically, so the 
    two parts of the equation produce the same result. 
 
\Paragraph{Total Expected Cost} 
The total expected operation cost,
    $\cost(\workload, \configuration)$, is computed by 
    weighing the empty point lookup cost $Z_0(\configuration)$ from 
    Equation~\eqref{eq:read}, the non-empty point lookup cost 
    $Z(\configuration)$ from Equation~\eqref{eq:read-non-empty}, the range 
    lookup cost $Q(\configuration)$ from Equation~\eqref{eq:range}, and the
    write cost $W(\configuration)$ from Equation~\eqref{eq:write} by their 
    proportion in the workload represented by the terms $z_0$, $z$, $q$ and $w$ 
    respectively (Note $z_0 + z_1 + q + w = 1$).  
This is the exact computation of the cost done in Equation~\eqref{eq:thecost}
    and in the definitions of the {\nominal} and {\robustw} problems
    (Equations~\eqref{eq:nominal_problem} and~\eqref{eq:robust_workload_problem}
    respectively).

\section{Uncertainty Benchmark}
\label{sec:uncertainty-benchmark}

In this section, we describe the uncertainty benchmark that we use to evaluate
    the robust tuning configurations given by \Endure, both analytically using
    the cost models, and empirically using RocksDB.
It consists of two primary components: (1) \emph{Expected workloads} and,
    (2) \emph{Benchmark set of sampled workloads}, described below.

\Paragraph{Expected Workloads}
We create robust tunings configurations for 15 expected workloads encompassing
    different proportions of query types.
We catalog them into \emph{uniform}, \emph{unimodal}, \emph{bimodal}, and
    \emph{trimodal} categories based upon the dominant query types.
While this breakdown of dominant queries is similar to benchmarks such as YCSB,
    we provide a more comprehensive coverage of potential workloads.
A minimum 1\% of each query type is always included in every expected workload
    to ensure a finite KL-divergence.
A complete list of all expected workloads is in Table~\ref{tab:expected-workloads}.

\begin{figure}
   \centering
   \includegraphics[width=\linewidth]{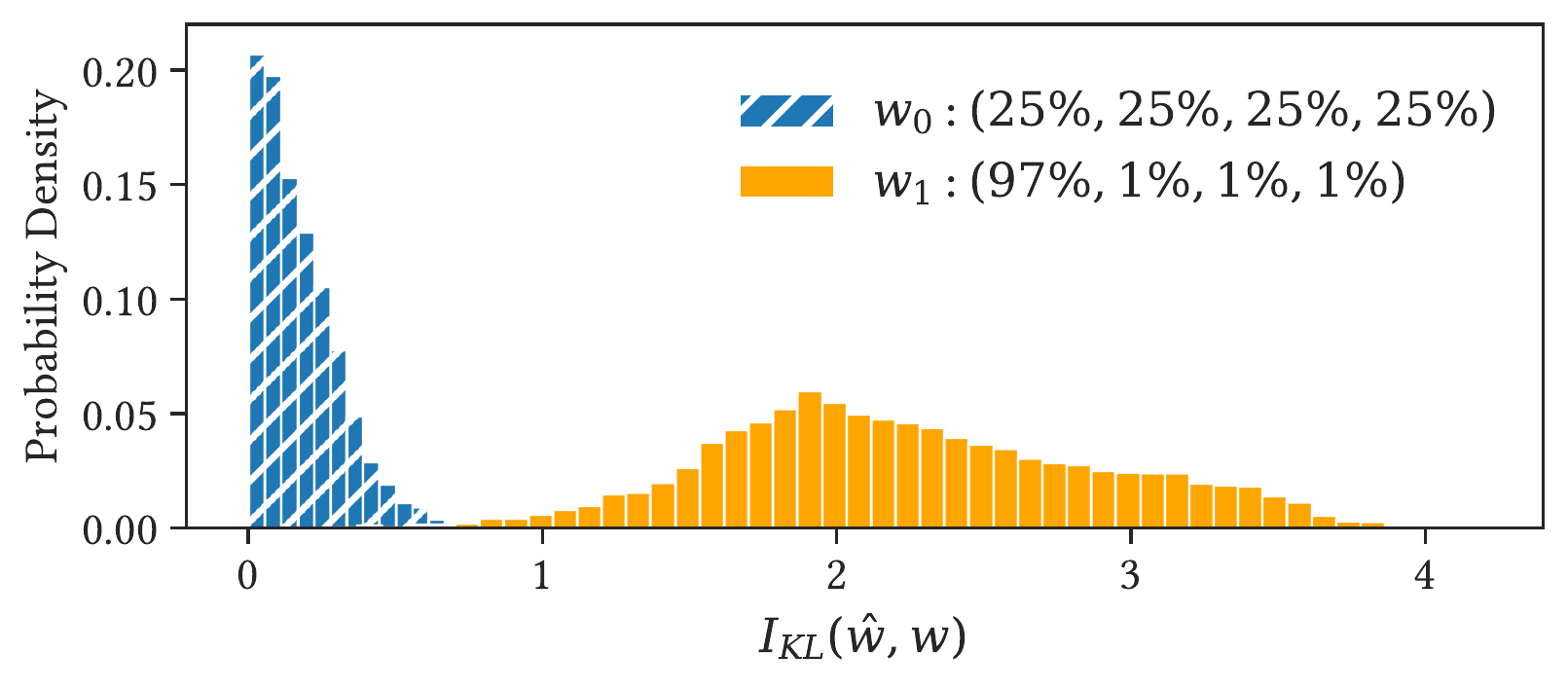}
   \caption{KL-divergence $I_{KL}(\hat{w}, w)$ histograms of the sampled
       workloads w.r.t. to expected workloads $w_0$ and $w_1$.}
   \label{fig:KL_divergence_histogram}
\end{figure}


\begin{table}
   \centering
   \begin{tabular}{c cccc l}
   \toprule
   Index & \multicolumn{4}{c}{$(\emptylookup, \nonemptylookup, \range, \update)$} & Type \\
   \toprule
   0 & 25\% & 25\% & 25\% & 25\% & \textbf{Uniform} \\
   \midrule
   1 & 97\% & 1\% & 1\% & 1\% & \textbf{Unimodal}\\
   2 & 1\% & 97\% & 1\% & 1\% & \\
   3 & 1\% & 1\% & 97\% & 1\% & \\
   4 & 1\% & 1\% & 1\% & 97\% & \\
   \midrule
   5 & 49\% & 49\% & 1\% & 1\% & \textbf{Bimodal}\\
   6 & 49\% & 1\% & 49\% & 1\% & \\
   7 & 49\% & 1\% & 1\% & 49\% & \\
   8 & 1\% & 49\% & 49\% & 1\% & \\
   9 & 1\% & 49\% & 1\% & 49\% & \\
   10 & 1\% & 1\% & 49\% & 49\% & \\
   \midrule
   11 & 33\% & 33\% & 33\% & 1\% & \textbf{Trimodal}\\
   12 & 33\% & 33\% & 1\% & 33\% & \\
   13 & 33\% & 1\% & 33\% & 33\% & \\
   14 & 1\% & 33\% & 33\% & 33\% & \\
   \bottomrule
   \end{tabular}
   \caption{Tested expected workloads.}
   \label{tab:expected-workloads}	
\end{table}

\Paragraph{Benchmark Set of Sampled Workloads}
We use the benchmark set of 10K workloads {\benchmark} as a \emph{test} dataset
    over which to evaluate the tuning configurations.
These configurations are generated as follows:
first, we independently sample the number of queries corresponding to each 
    query type uniformly at random from a range $(0, 10000)$ to obtain a 
    $4$-tuple of query counts.
Next, we divide the individual query counts by the total number of queries in 
    the tuple to obtain a random workload that is added to the benchmark set.
We use the actual query counts during the system experimentation where we
    execute individual queries on the database.

This type of workload breakdown can commonly be seen in LSM trees as shown in
    a survey of workloads in Facebook's own pipeline~\cite{Cao2020}.
The authors report that ZippyDB, a distributed KV store that uses RocksDB,
    experiences workloads with 78\% gets, 19\% writes, and 3\% range reads.
This breakdown is similar to workload $11$, and the exact workload is in
    the benchmark set {\benchmark}.

Note that while the same {\benchmark}
    is used to evaluate different tunings, it represents a different 
    distribution of KL-divergences for the corresponding expected workload 
    associated with each tuning.
As an example, in Figure \ref{fig:KL_divergence_histogram}, we plot the 
    distribution of KL-divergences of sampled workloads in {\benchmark} 
    w.r.t. the expected workloads $\workload_0$ and $\workload_1$ from 
    Table \ref{tab:expected-workloads}.
In the next two sections, we use our uncertainty benchmark to show that tuning
    with {\Endure} achieves significant performance improvement using both a
    model-based analysis (Section~\ref{sec:model-evaluation}), and an experimental
    study (Section~\ref{sec:system-evaluation}).

\section{Model-Based Evaluation}
\label{sec:model-evaluation}

We now present our detailed model-based study of {\Endure} that uses more than
    10000 different noisy workloads for all 15 expected workloads, showing
    performance benefit of up to $5\times$. 
In addition, we show that {\Endure} perfectly matches the classical tuning when
    there is no uncertainty, that is when the observed workload always matches
    the expected one, and we pass this information to the robust tuner.
Further, we provide recommendations on how to choose uncertainty parameters. 

\subsection{Evaluation Metrics}

\Paragraph{Normalized Delta Throughput ($\Delta$)} Defining throughput as the 
    reciprocal of the cost of executing a workload, we measure the
    normalized delta throughput of a configuration {$\configuration_2$}
    w.r.t. another configuration {$\configuration_1$} for a given 
    workload {\workload} as follows, 
    \[
        \Delta_{\workload}(\configuration_1, \configuration_2) = 
        \frac{\nicefrac{1}{C(\workload, \configuration_2)} -
            \nicefrac{1}{C(\workload, \configuration_1)}}
            {\nicefrac{1}{C(\workload, \configuration_1)}}.
    \]
$\Delta_{\workload}(\configuration_1, \configuration_2) > 0$ implies that
    {$\configuration_2$} outperforms {$\configuration_1$} when executing
    a workload {\workload} and vice versa when
    $\Delta_{\workload}(\configuration_1, \configuration_2) < 0$.

\Paragraph{Throughput Range ($\Theta$)} While normalized delta throughput 
    compares two different tunings, we use the throughput
    range to evaluate an individual tuning ${\configuration}$ 
    w.r.t. the  benchmark set {\benchmark} as follows,
    \[
        \Theta_{\mathcal{B}}(\configuration) = 
        \max_{\workload_0, \workload_1 \in \mathcal{B}}
            \bigg(\frac{1}{C(\workload_0, \configuration)} 
            - \frac{1}{C(\workload_1, \configuration)}\bigg).
    \]
$\Theta_{\benchmark}(\configuration)$ intuitively captures the best
    and the worst-case outcomes of the tuning {\configuration}.
A smaller value of this metric implies higher consistency in performance.

\begin{figure}
    \centering
    \includegraphics[width=\linewidth]{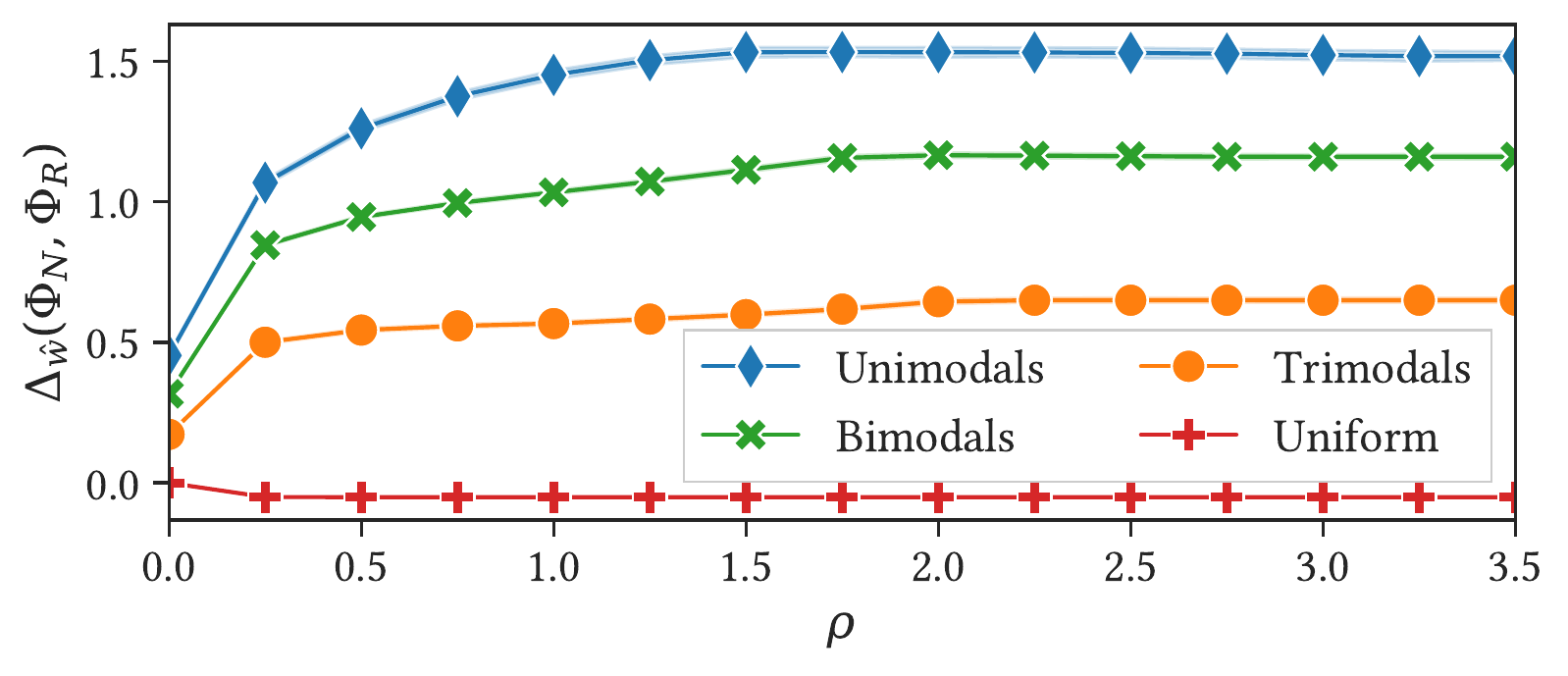}
    \caption{Average delta throughput $\Delta_{\hat{w}}(\Phi_{N}, \Phi_{R})$
        for each category of expected workload.
    }
    \label{fig:delta_throughput_workload_type}
\end{figure}

\begin{figure*}
    \centering
    \includegraphics[width=\textwidth]{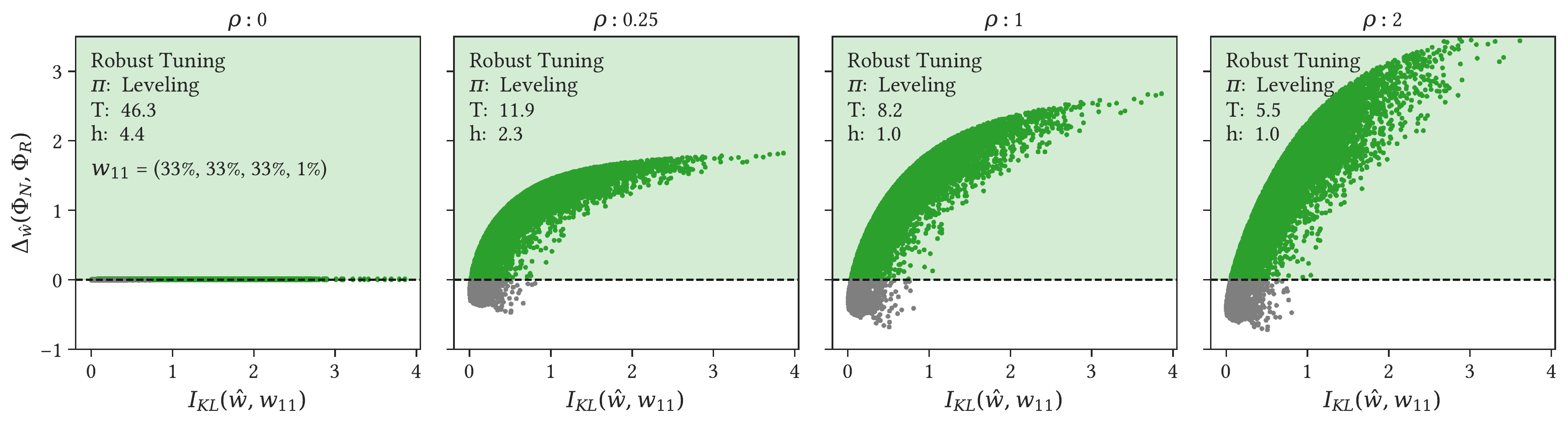}
    \caption{Impact of $\rho$ on normalized delta throughput 
        $\Delta_{\hat{w}]}(\Phi_{N}, \Phi_{R})$ for tunings with expected
        workload $w_{11}$.
    }
    \label{fig:scatterplot_evolution_rho}
\end{figure*}

\begin{figure*}[tbp]
    \centering
    \captionsetup{aboveskip=0.5em}
    \begin{subfigure}[h]{0.74\textwidth}
        \includegraphics[scale=0.5]{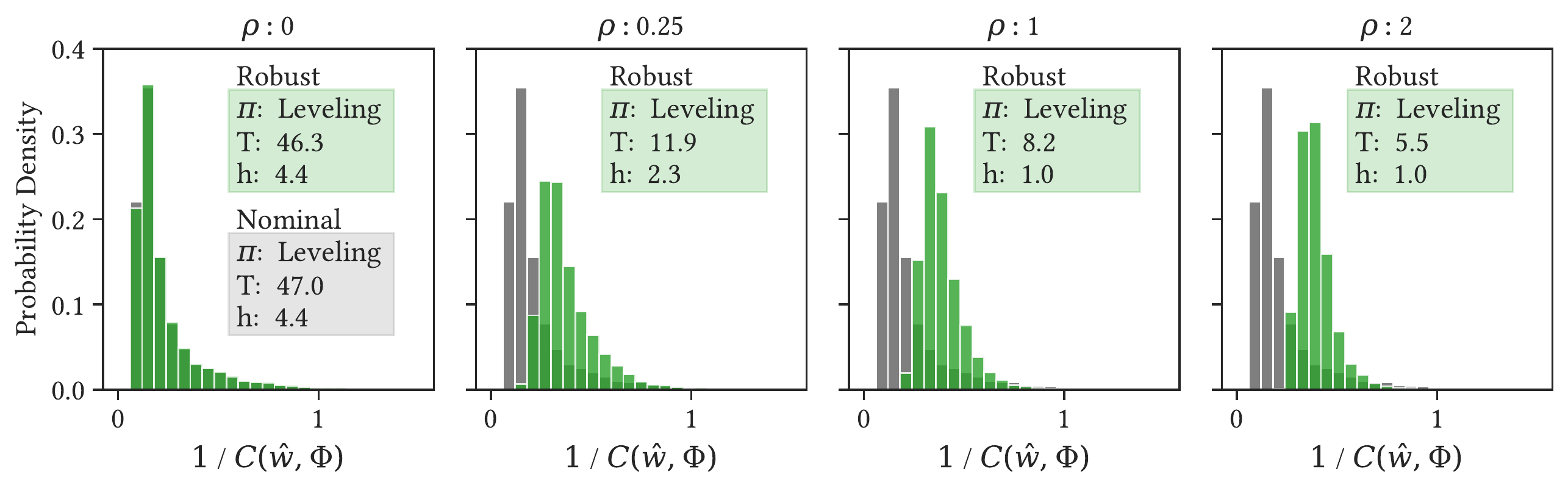}
        \caption{Histograms of throughput $1 / C(\hat{w}, \Phi)$ for
            tunings with expected workload $w_{11}$ }
        \label{fig:overlapping_histogram}
    \end{subfigure}
    \begin{subfigure}[h]{0.24\textwidth} 
        \includegraphics[scale=0.5]{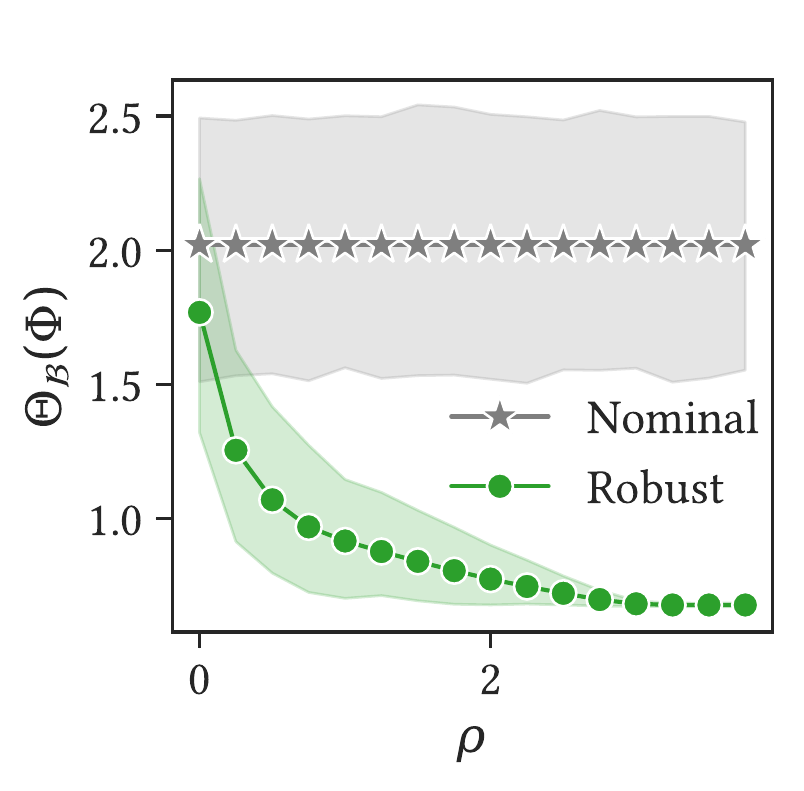}
        \caption{Throughput range $\Theta_{\mathcal{B}}(\Phi)$}
        \label{fig:throughput_range_evolution}
    \end{subfigure}
    \caption{Impact of $\rho$ on throughput.}
    \label{fig:rho_throughput_range_impact}
\end{figure*}

\subsection{Experiment Design}
To evaluate the performance of our proposed robust tuning approach, 
    we design a large-scale experiment comparing different tunings
    over the sampled workloads in {\benchmark} using the analytical cost
    model.
For each of the expected workloads in Table \ref{tab:expected-workloads},
    we obtain a single nominal tuning configuration (${\configuration_N}$) by 
    solving the {\nominal} problem. 
For 15 different values of $\rho$ in the range (0.0, 4.0) with a step size of
    0.25, we obtain a set of robust tuning configurations ($\configuration_R$)
    by solving the {\robustw} problem.
Finally, we individually compare each of the robust tunings with the nominal
    over the 10,000 workloads in {\benchmark} to obtain over 2 million
    comparisons.
While computing the costs, we assume that the database contains 10 million
    entries each of size 1 KB. 
The analysis presented in the following sections assumes a total available memory of
    10 GB.
In the following sections, for brevity purposes, we present representative 
    results corresponding to individual expected workloads and specific system 
    parameters.
However, we exhaustively confirmed that changing these parameters does not 
    qualitatively affect the outcomes of our experiment.

\subsection{Results}
\label{sec:model-results}
Here, we present an analysis of the comparisons between the robust and 
    the nominal tuning configurations.
Using an off-the-shelf global minimizer from the popular Python optimization
    library SciPy \cite{2020SciPy-NMeth}, we obtain both nominal and robust
    tunings with the runtime for the above experiment being less than 10
    minutes.

\Paragraph{Comparison of Tunings}
First, we address the question -- \textit{is it beneficial to adopt robust 
    tunings relative to the nominal tunings?}
Intuitively, it should be clear that the performance of nominally tuned 
    databases would degrade when the workloads being executed on the database 
    are significantly different from the expected workloads used for tuning.
In Figure~\ref{fig:delta_throughput_workload_type},
    we present performance comparisons between the robust and the nominal 
    tunings for different values of uncertainty parameter $\rho$.
We observe that robust tunings provide substantial benefit in terms of 
    normalized delta throughput for \emph{unimodal}, \emph{bimodal}, and 
    \emph{trimodal} workloads.
The normalized delta throughput 
    $\Delta_\obsworkload(\configuration_N, \configuration_R)$
    shows over 95\% improvement on average over all $\obsworkload \in \benchmark$ 
    for robust tunings with $\rho \geq 0.5$, 
    when the expected workload used during tuning belongs to one of these 
    categories.
For \emph{uniform} expected workload, we observe that the nominal tuning 
    outperforms the robust tuning by a modest 5\%.

Intuitively, \emph{unbalanced} workloads result in overfit nominal tunings.
Hence, even small variations in the observed workload can lead to 
    significant degradation in the throughput of such nominally tuned databases. 
On the other hand, robust tunings by their very nature take into account such
    variations and comprehensively outperform the nominal tunings.
In the case of the uniform expected workload $\workload_0$, 
    Figure ~\ref{fig:KL_divergence_histogram} shows us that instances of 
    high values of KL-divergence are extremely rare.
In this case, when tuned for high values of $\rho$, the robust tunings are 
    unrealistically pessimistic and lose out some performance relative to the 
    nominal tuning.

\Paragraph{Impact of Tuning Parameter $\rho$}
Next, we address the question -- \emph{how does the uncertainty tuning parameter
    $\rho$ impact the performance of the robust tunings?}
In Figure~\ref{fig:scatterplot_evolution_rho}, we take a deep dive into
    the performance of robust tunings for an individual
    expected workload for different values of $\rho$.
We observe that the robust tunings for $\rho=0$ i.e., zero uncertainty, are 
    very close to the nominal tunings.
As the value of $\rho$ increases,
    its performance advantage over the nominal tuning for the observed 
    workloads with higher KL-divergence w.r.t. expected workload
    increases.
Furthermore, the robustness of such configurations have logically sound 
    explanations.
The expected workload in 
    Figure~\ref{fig:scatterplot_evolution_rho} consists of just 1\% writes.
Hence, for low values of $\rho$, the robust tuning has higher size-ratio leading 
    to shallower LSM trees to achieve good read performance.
For higher values of $\rho$, the robust tunings anticipate an increasing
    percentage of write queries and hence limit the size-ratio to achieve 
    higher throughput.

In Figure~\ref{fig:rho_throughput_range_impact}, we show the impact of tuning
    parameter $\rho$ on the throughput range.
In Figure~\ref{fig:overlapping_histogram} we plot a histogram of 
    the nominal and robust throughputs for workload $\workload_{11}$. 
As the value of $\rho$ increases, the interval size between the lowest and the 
    highest throughputs for the robust tunings consistently decreases.
We provide further evidence of this phenomenon in 
    Figure~\ref{fig:throughput_range_evolution}, by plotting the decreasing  
    throughput range $\Theta_\benchmark(\configuration_R)$  
    averaged across all the expected workloads. 
Thus, robust tunings not only provide a higher average throughput over all 
    $\obsworkload \in \benchmark$, but they have a more consistent
    performance (lower variance) compared to the nominal tunings.

\begin{figure}[t]
    \centering
    \includegraphics[scale=0.25]{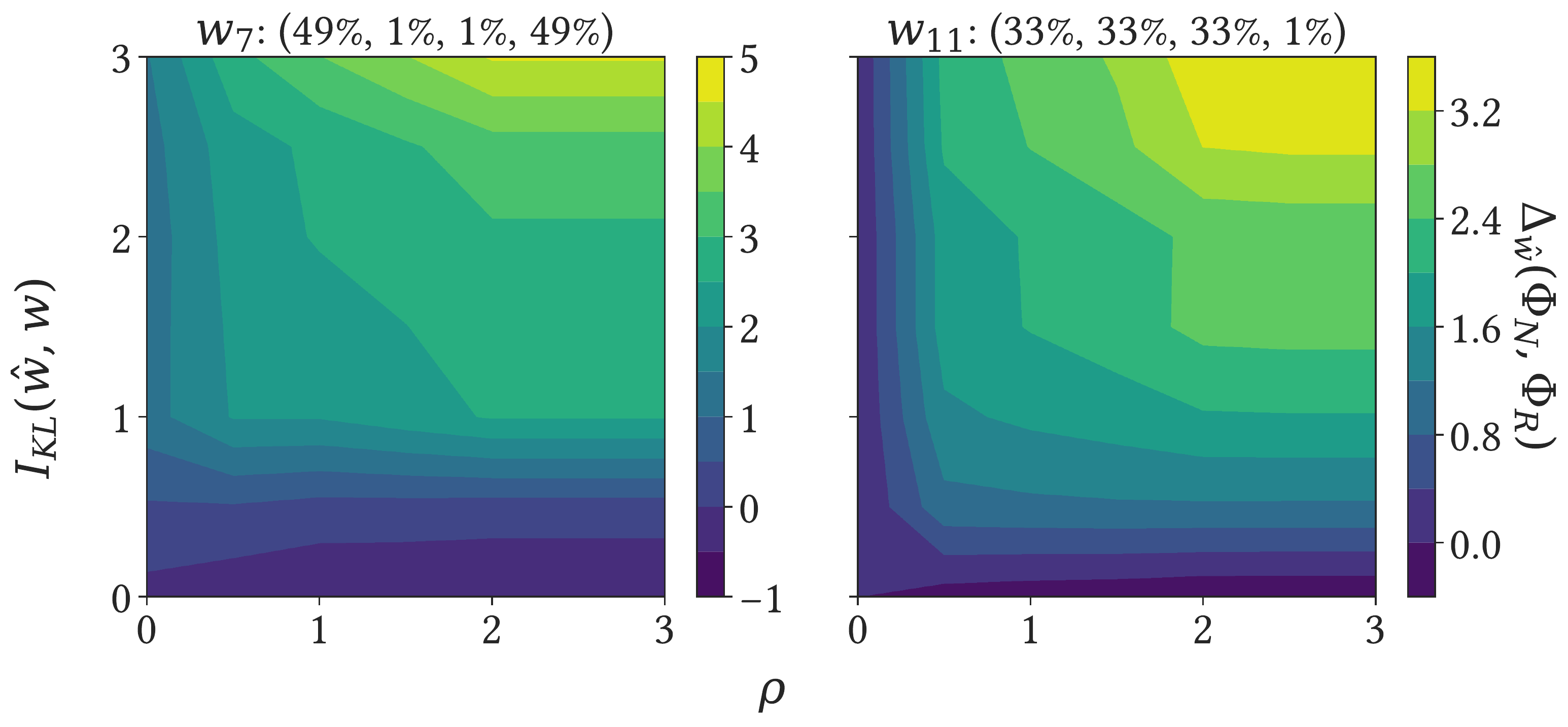}
    \caption{Delta throughputs $\Delta_{\hat{w}}(\Phi_{N},
        \Phi_{R})$ for $\rho$ vs $I_{KL}(\hat{w}, w)$.}
    \label{fig:rho_vs_rho_hat}
\end{figure}

\Paragraph{How to Choose $\rho$} Now, we address the question -- 
\emph{What is the appropriate choice for the value of uncertainty parameter $\rho$?}
We provide guidance on the choice of $\rho$ in absence of perfect knowledge
    regarding the future workloads that are likely to be executed on the
    database.
Intuitively, we expect the robust tunings to be only weak when they are tuned 
    for either too little or too much uncertainty.
In Figure~\ref{fig:rho_vs_rho_hat}, we explore the relationship between $\rho$
and the KL-divergence $I_{KL}(\obsworkload, \workload)$ for 
    $\obsworkload \in \benchmark$, 
    by making a contour plot of the corresponding normalized delta throughput
    $\Delta_{\obsworkload}(\configuration_N, \configuration_R)$.
We confirm our intuition that nominal tunings compare favorably with our
    proposed robust tunings only in two scenarios viz., (1) when observed 
    workloads are extremely similar to the expected workload (close to zero 
    observed uncertainty), and (2) when the robust tunings assume extremely low 
    uncertainty with $\rho < 0.2$ while the observed variation is higher.
Based on this evidence, we can advise a potential database administrator that
    mean KL-divergences between pairs of historically observed workloads would
    be a reasonable value of $\rho$ while deploying robust tunings in practice.

\section{System-Based Evaluation}
\label{sec:system-evaluation}

In this section, we deploy {\Endure} as the tuner of the state-of-the-art
    LSM-based engine RocksDB, and we show that RocksDB achieves up to 90\% lower
    workload latency in the presence of uncertainty.
We further show that the tuning cost is negligible, and the effectiveness of
    {\Endure} is not affected by data size.

\subsection{Experimental Setup \& Measurements}

Our server is configured with two Intel Xeon Gold 6230 processors, 384 GB 
    of main memory, a 1 TB Dell P4510 NVMe drive, CentOS 7.9.2009, and a default
    page size of 4 KB.
We use Facebook's RocksDB, a popular LSM tree-based storage system, to evaluate
    our approach~\cite{FacebookMyRocks}.
While RocksDB provides implementations of leveling and tiering policies, the
    system implements micro-optimizations not common across all LSM
    tree-based storage engines.
Therefore, we use RocksDB's event hooks to implement both classic leveling and
    tiering policies to benchmark the common compaction
    strategies~\cite{RocksDB2020a}.
Following the Monkey memory allocation scheme~\cite{Dayan2017}, we allocate 
    different bits per element for Bloom filters per level. 
To obtain an accurate count of block accesses we enable direct I/Os for both
    queries and compaction and disable the block cache.
The remaining parameters such as buffer size are set by the tuning.

\Paragraph{Empirical Measurements}
We use the internal RocksDB statistics module to measure the number of logical
    block accesses during reads, bytes flushed during writes, and bytes read and
    written in compactions.
The number of logical blocks accessed during writes is calculated by 
    dividing the number of bytes reported by the default page size.
To estimate the amortized cost of writes, we compute the I/Os from compactions 
    across all workloads of a session and redistribute them across write queries.
Our approach of measuring average I/Os per query allows us to compare the
    effects of different tuning configurations, while simultaneously minimizing
    the effects of extraneous factors on the database performance.

\begin{figure*}[t]
    \centering
    \includegraphics[width=\linewidth]{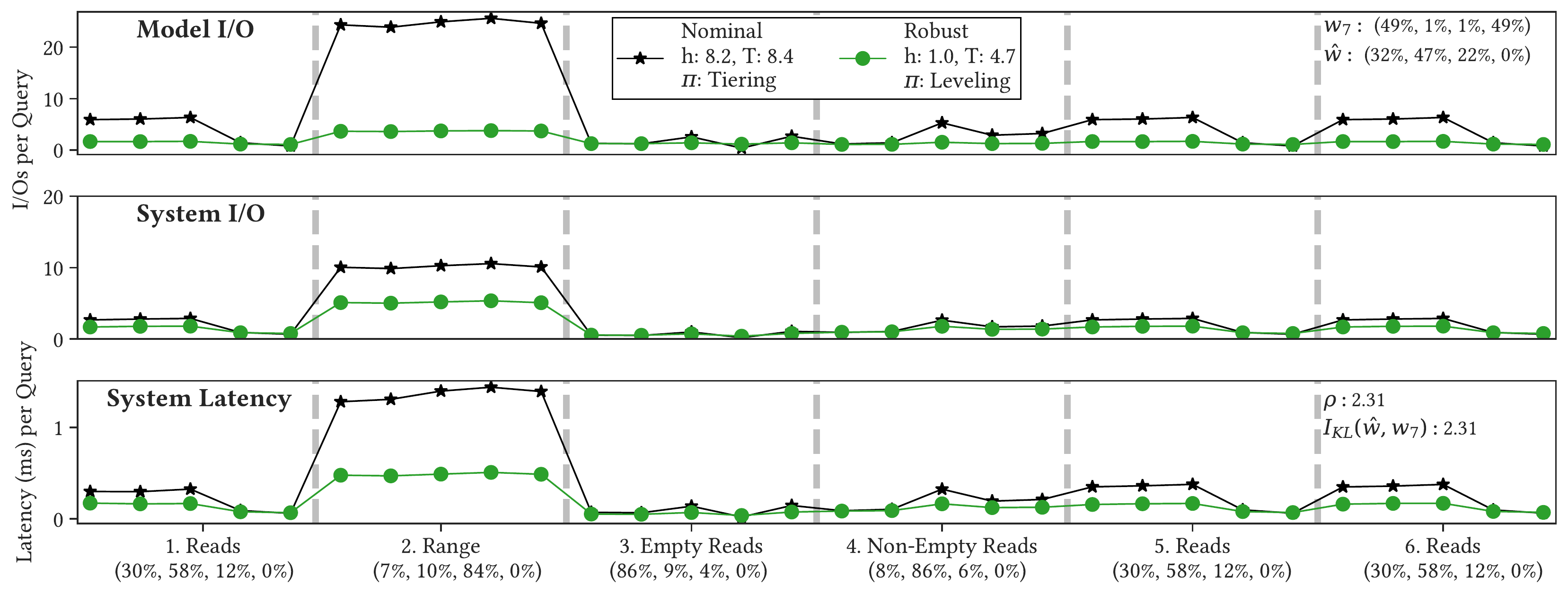}
    \caption{System and model performance for robust and nominal
        tunings in a read-only query sequence. Here the tuning parameter $\rho$
        matches the observed value of $I_{KL}(\hat{w}, w_{7})$. Each session
        contains the label and average workload.
    }
    \label{fig:query_seq_reads_1}
\end{figure*}

\begin{figure*}[t]
    \centering
    \includegraphics[width=\linewidth]{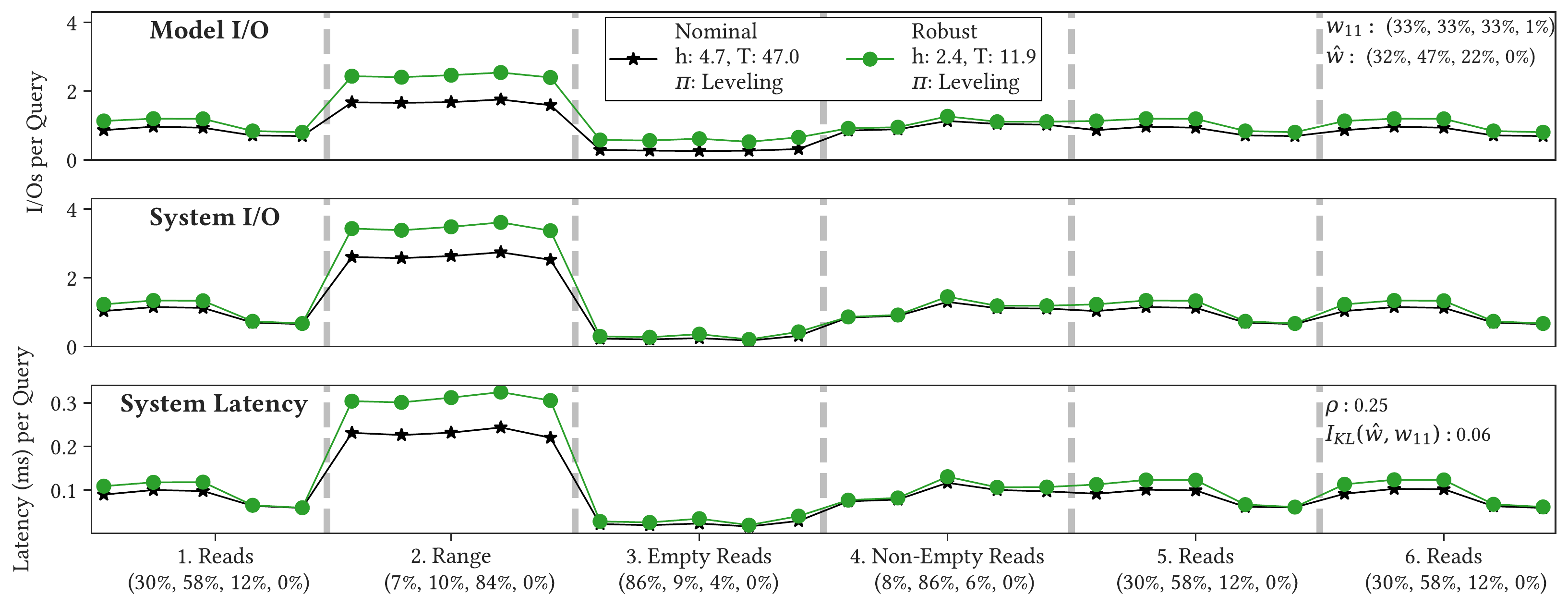}
    \caption{Read-only sequence where the observed workloads {\obsworkload}
        is close to the expected, hence $\rho$ and $I_{KL}(\hat{w},
        w_{11})$ deviate.
    }
    \label{fig:query_seq_reads_2}
\end{figure*}

\subsection{Experiment Design}

To evaluate the performance of our proposed robust tuning approach, we create
    multiple instances of RocksDB using different tunings and empirically
    measure their performance by executing workloads from the uncertainty
    benchmark {\benchmark}.
To measure the steady-state performance of the database, each instantiation is
    initially bulk loaded with the exact same sequence of 10 million
    unique key-value pairs each of size 1 KB.
Each key-value entry has a 16-bit uniformly at random 
    sampled key, with the remaining bits being allocated to a randomly 
    generated value.

While evaluating the performance of the database, we sample a sequence of 
    workloads from the benchmark set {\benchmark}.
Each sequence is cataloged into one of the categories — \emph{expected}, 
    \emph{empty read}, \emph{non-empty read}, \emph{read}, \emph{range}, 
    and \emph{write} — based on the dominant query type in the workloads
    in the sequence.
Specifically, the \emph{expected} session contains workloads with a
    KL-divergence less than 0.2 w.r.t. the expected workload used for tuning. 
In all other sessions, the dominant query type encompasses 80\% of the total
    queries in the session.
The remaining 20\% of queries may belong to any of the query types.
While generating keys of the queries to run on the database, we ensure that
    non-empty point reads query a key that exists in the database, while the
    empty point reads query a key that is not present in the database but is
    sampled from the same domain.
All range queries are generated with minimal selectivity $S_{RQ}$ to act as 
    short range queries reading on average zero to two pages per level.
Lastly, write queries consist of randomly generated keys that are guaranteed to
    be unique from the existing keys in the database.
Initializing RocksDB and bulk loading requires 30 minutes, while
    execution of individual workloads are 5 minutes on average.    

\begin{figure*}[t]
    \centering
    \includegraphics[width=\linewidth]{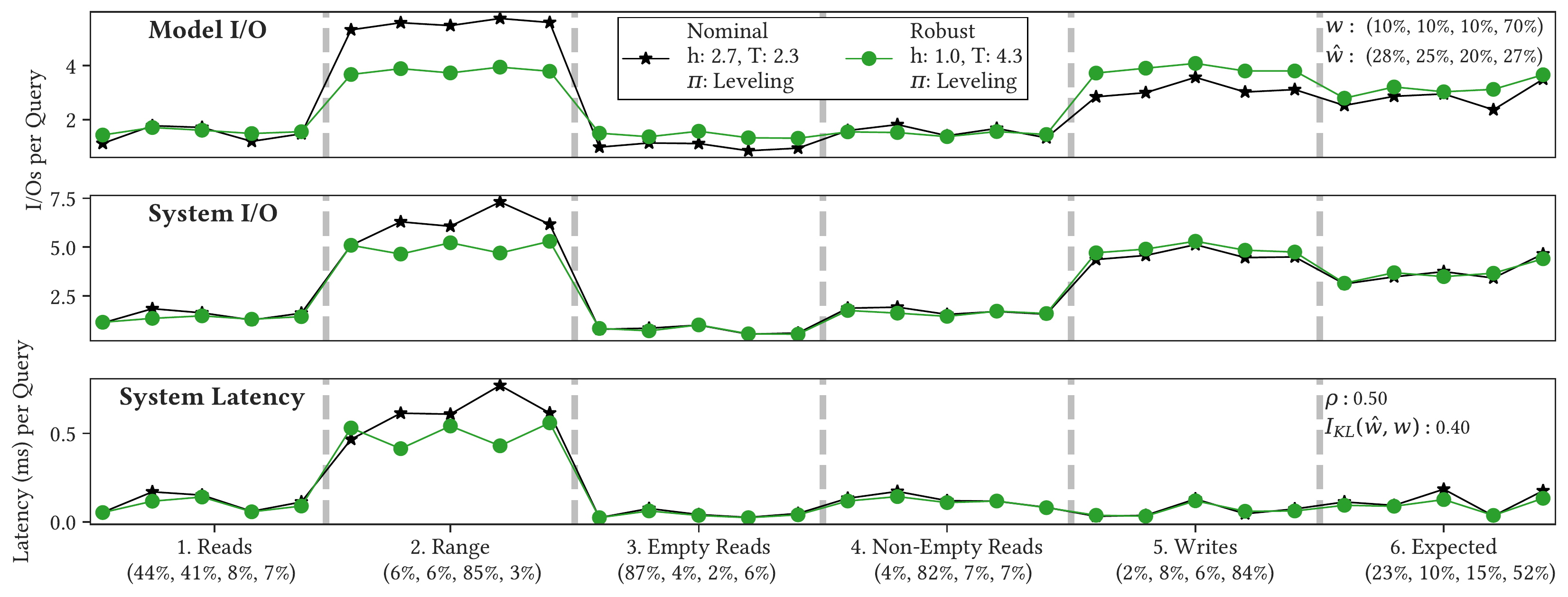}
    \caption{Sequence where $\rho$ and $I_{KL}(\hat{w}, w)$ closely
        match. Performance flucuates with writes as it changes the tree
        structure.}
    \label{fig:query_seq_hybrid_1}
\end{figure*}

\begin{figure*}[t]
    \centering
    \includegraphics[width=\linewidth]{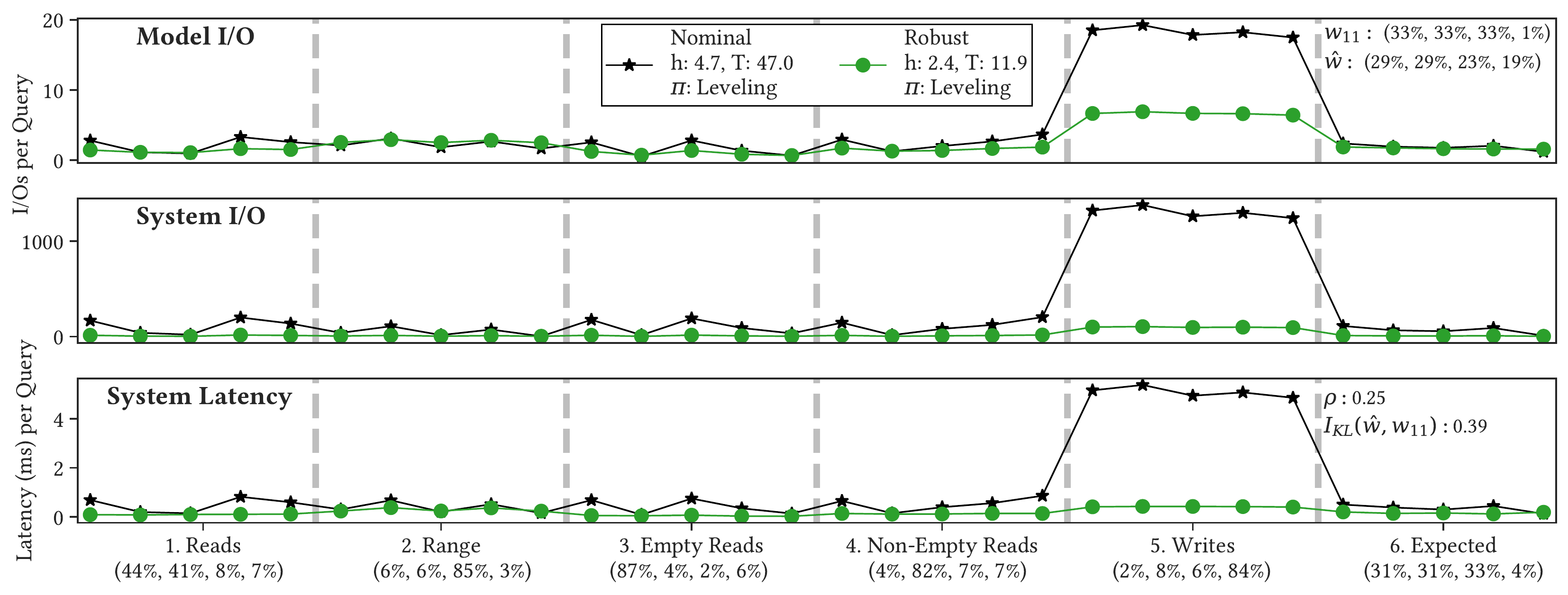}
    \caption{Sequence where $\rho$ and $I_{KL}(\hat{w}, w_{11})$ closely
        match. Both system I/O and latency show reductions of up to $90\%$.}
    \label{fig:query_seq_hybrid_2}
\end{figure*}

\subsection{Experimental Results}
\label{sec:system-results}

In this section, we replicate key insights from
    Section~\ref{sec:model-evaluation}, evaluate system performance, and show that
    {\Endure} scales with database size.
Due to space constraints, we present results for 2 representative expected
    workloads, $\workload_{7}$ and $\workload_{11}$.
An interested reader can find results for all workloads in an extended version
    of this work \cite{Huynh2021}.

\Paragraph{Cost of Tuning}
For every experiment, we perform either the nominal or the robust tuning prior to
    the experiments.
Both nominal and robust tuning takes less than 10ms, which is negligible  w.r.t.
    the workload execution time. 

\Paragraph{Read Performance}
We begin by examining the system performance and verifying that the
    model-predicted I/O and the system-measured I/O match when considering read
    queries in Figures~\ref{fig:query_seq_reads_1}
    and~\ref{fig:query_seq_reads_2}.
In both figures, we include the model-predicted I/Os per query (top), the
    I/Os per query measured on the system (middle), and the system latency
    (bottom), for both nominal and robust tunings across different read
    sessions.
The empirical measurements confirm the cost model predictions and show that the
    predicted performance benefits from the model translate to similar
    performance benefits in practice.
Note that the discrepancy observed between the relative performance between the
    nominal and the robust tunings in the presence of range queries (session 2
    in Figure~\ref{fig:query_seq_reads_1}) is due to the fence pointers in
    RocksDB.
The analytical model does not account for fence pointers allowing the system to
    completely skip a run, which may reduce the measured I/Os for short range
    queries compared to the predicted I/Os.

\Paragraph{Write Performance}
In presence of writes (Figures~\ref{fig:query_seq_hybrid_1}
    and~\ref{fig:query_seq_hybrid_2}),
the model is still predicting the disk accesses successfully and {\Endure}
leads to significant performance benefits as expected from the 
model-based analysis.
Note that now the structure of the LSM tree is continually changing across all
    sessions due to the compactions caused by write queries.
For example, the dips in measured I/Os in the range query session in
    Figure~\ref{fig:query_seq_hybrid_1} are the result of compactions triggered
    by write queries in preceding workloads leading to empty levels.
Additionally, writes may appear instantaneous w.r.t. system latency as seen in
    Figure~\ref{fig:query_seq_hybrid_1} due to RocksDB assigning compactions to
    background threads.
For Figure~\ref{fig:query_seq_hybrid_1}, we see that the particular tuning causes
    compactions to be fast enough and not stall any future queries. 
Thus, RocksDB does not experience any latency disruptions in performance. 
In contrast, in Figure~\ref{fig:query_seq_hybrid_2} we see that the nominal tuning
    causes compactions to be costly as the large size-ratio {\sizeratio} leads
    to a shallow tree with extremely large levels. 
Thus, a compaction occurring in the write query dominated session triggers a 
    sort at the lower levels of the tree resulting in a higher number of I/Os 
    than predicted by the model.
As a result, we observe that robust tuning reduces I/O and latency by up to
    $90\%$.
Overall, Figures~\ref{fig:query_seq_reads_1}--\ref{fig:query_seq_hybrid_2} 
    confirm that our analytical model can accurately capture the relative 
    performance of different tunings.

\begin{figure*}
    \centering
    \includegraphics[width=\linewidth]{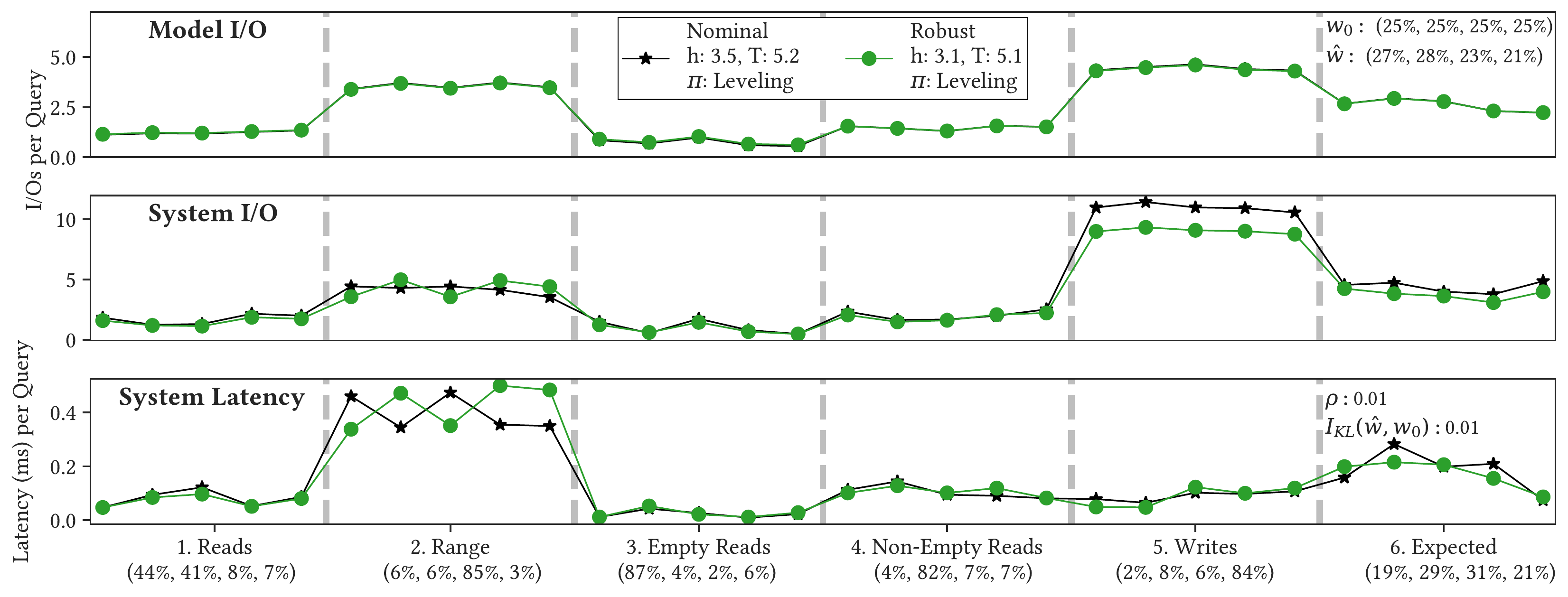}
    \caption{Uniform workload $w_{0}$.}
    \label{fig:uniform_wl}
\end{figure*}

\Paragraph{Uniform Workload}
In Figure~\ref{fig:uniform_wl} we show an instance where the expected workload
    and the observed workload are similar ($I_{KL}(\obsworkload, \workload_{0})
    = 0.01$), and we assign the correct $\rho$.
Note that both the robust and nominal tuning produce similar designs.
This similarly is reflected in performance, where we see both tunings
    producing similar system I/O and latency measurements.
The only different appears in the writes, session 5, where we observe the nominal
    tuning causes system to issue more I/Os.
Because the nominal design issues more bits per elements to the bloom filter, 
    the memory buffer is smaller implying more compactions are triggered.

\begin{figure*}
    \centering
    \includegraphics[width=\linewidth]{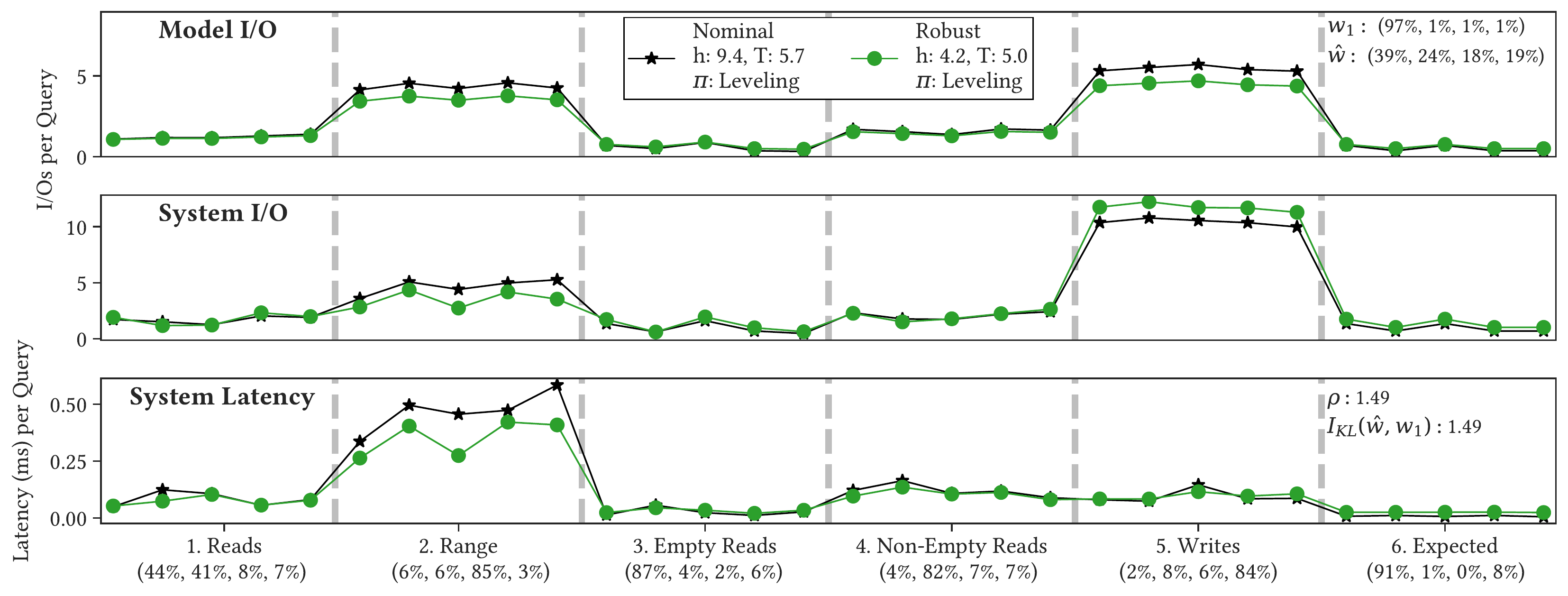}
    \includegraphics[width=\linewidth]{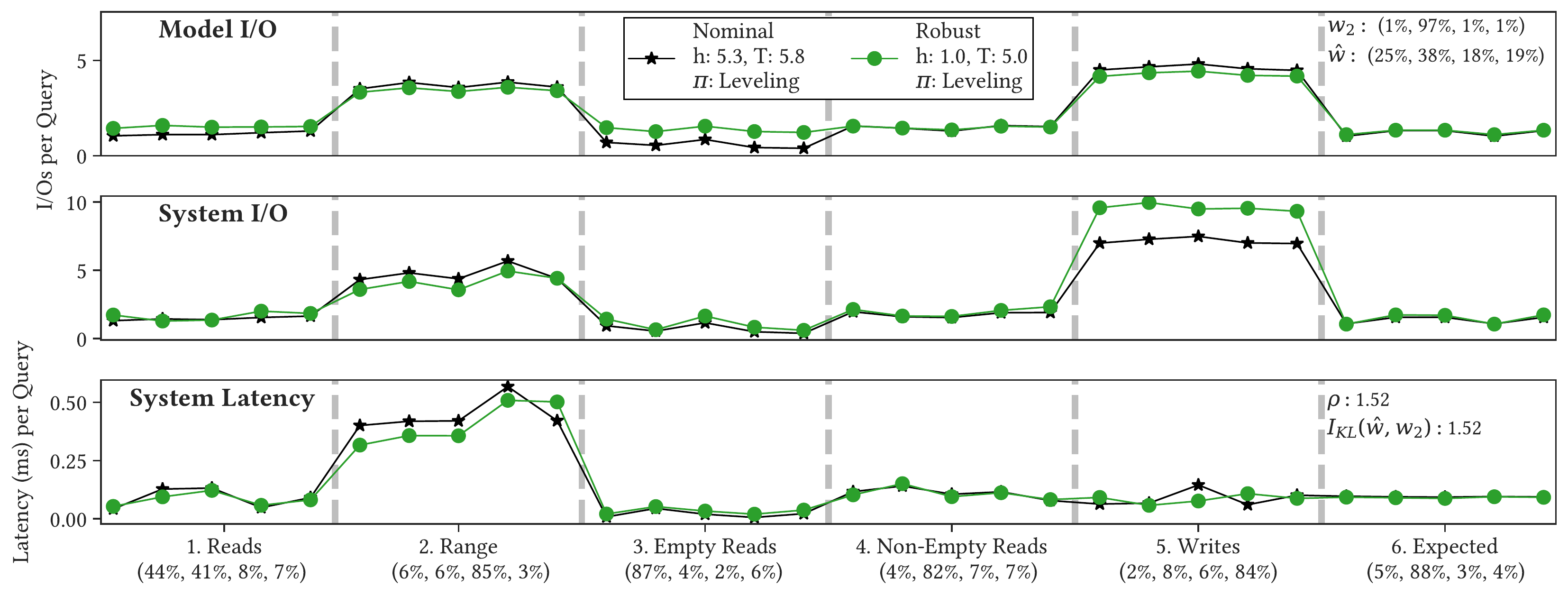}
    \caption{Unimodal workloads $w_{1}$ and $w_{2}$.}
    \label{fig:unimodal_wl_12}
\end{figure*}

\begin{figure*}
    \centering
    \includegraphics[width=\linewidth]{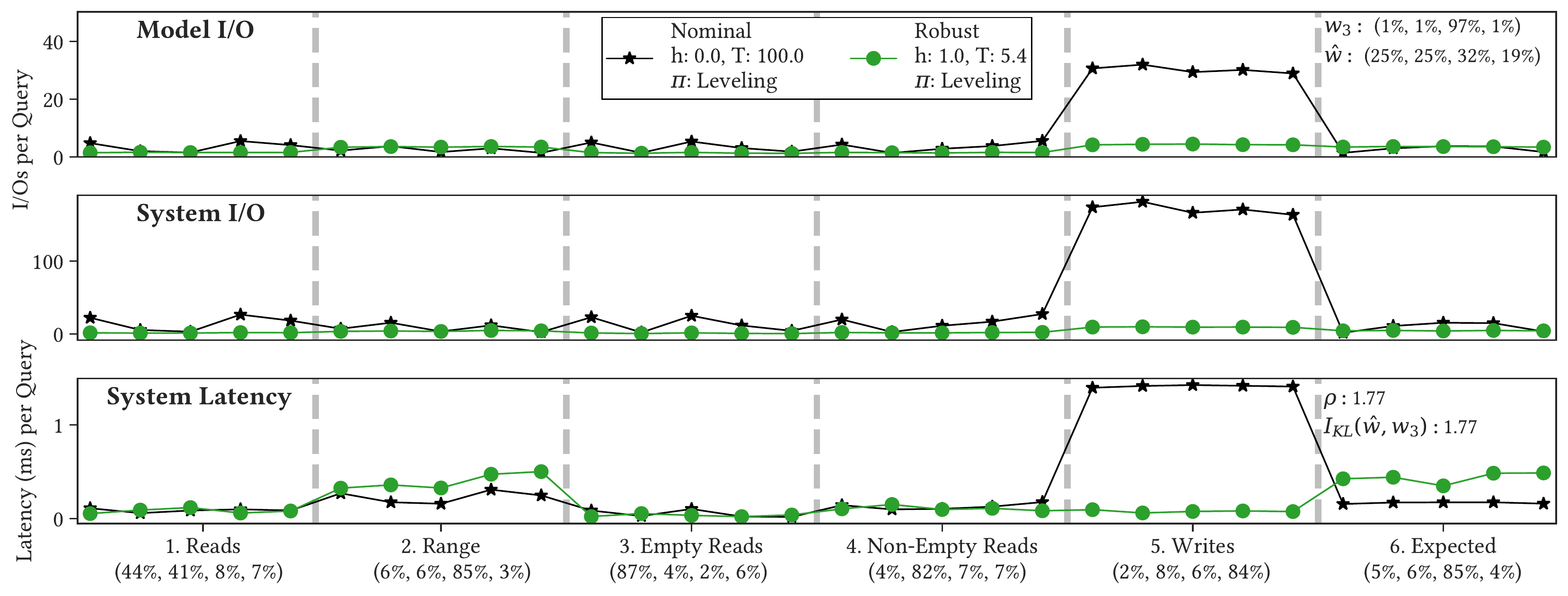}
    \includegraphics[width=\linewidth]{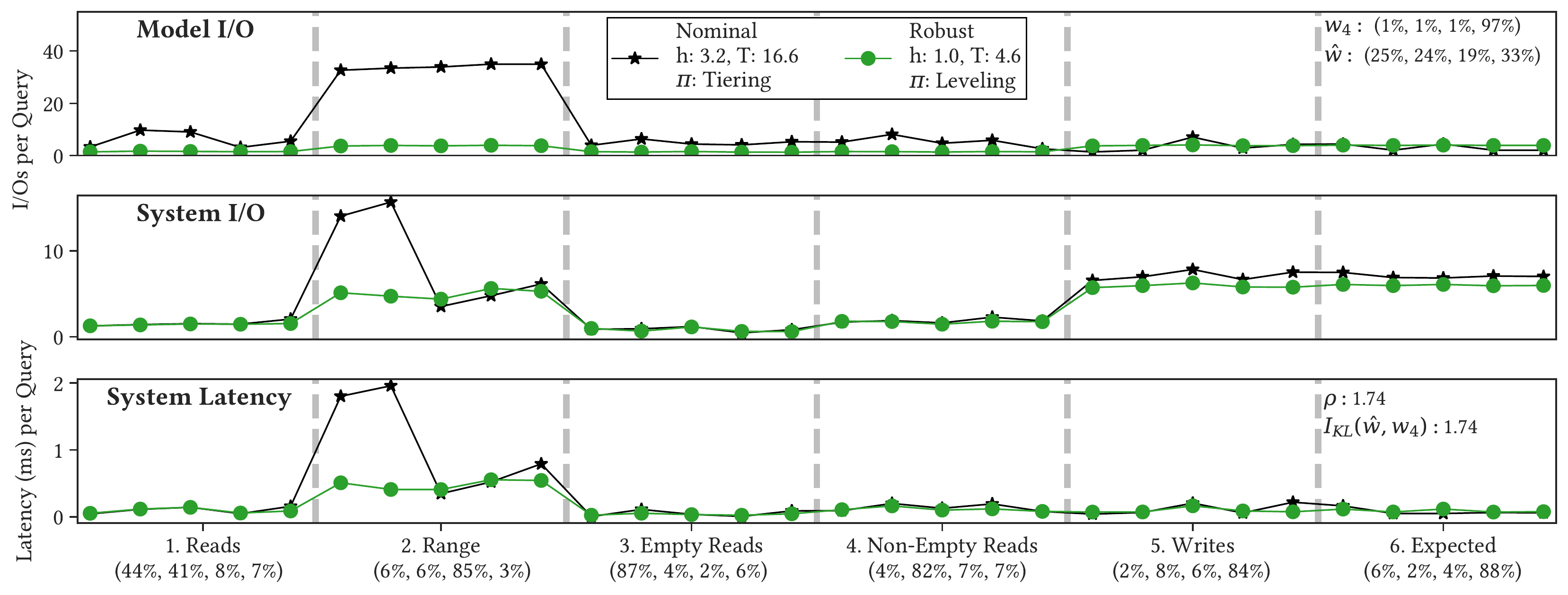}
    \caption{Unimodal workloads $w_{3}$ and $w_{4}$.}
    \label{fig:unimodal_wl_34}
\end{figure*}

\Paragraph{Unimodal Workloads}
In Figures~\ref{fig:unimodal_wl_12} and~\ref{fig:unimodal_wl_34} we show results
    for the unimodal workloads.
Note that, similar to the previous graphs, when introducing writes into the
    workload the latency may appear instantaneous as compactions are issued on
    background threads.
We see for $\workload_{1}$ the system I/O and model I/O match up well, however, the
    change in trend for the write session occurs due to the rounding of size
    ratio.
Classical LSM trees cannot have fractional size ratios, therefore we round up
    when deploying tunings onto the physical system.
For $\workload_{2}$ we see a similar trend.
In $\workload_{3}$ we observe that the nominal assigns a large size ratio due to the
    expectation of range queries being the majority query type.
Once a heavy write session occurs, we see the tuning performing much worse than
    the nominal as the compaction costs becomes much more expensive as your size
    ratio increases.
In contrast, we see the nominal performing well in range queries w.r.t.
    system latency, however, this benefit does not offset the consistency for
    query response time we see from the robust tuning.
Lastly in $\workload_{4}$ we see that because the tree structure changes over
    time in the presence of writes, there's a dip in measured system I/O and
    latency.
Additionally, we see the baseline for system I/O per query increasing post
    session 5 as the tree changes shape.

\begin{figure*}
    \centering
    \includegraphics[width=\linewidth]{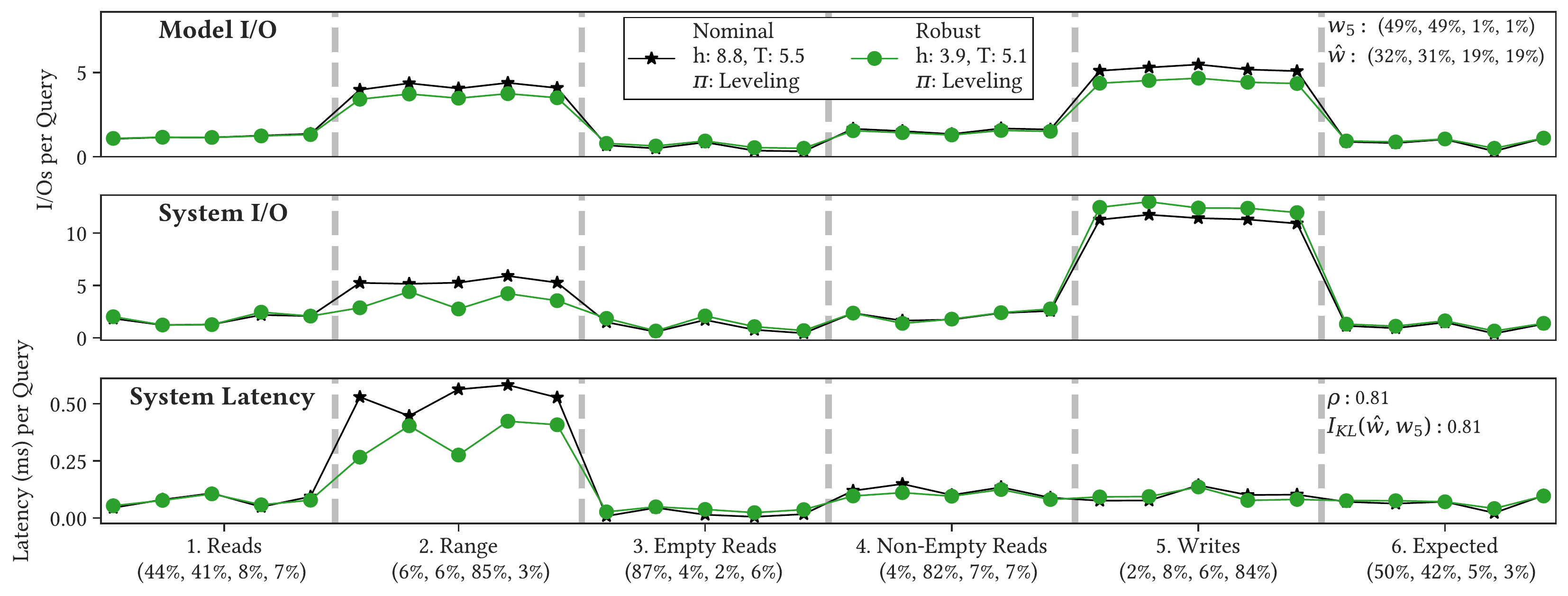}
    \includegraphics[width=\linewidth]{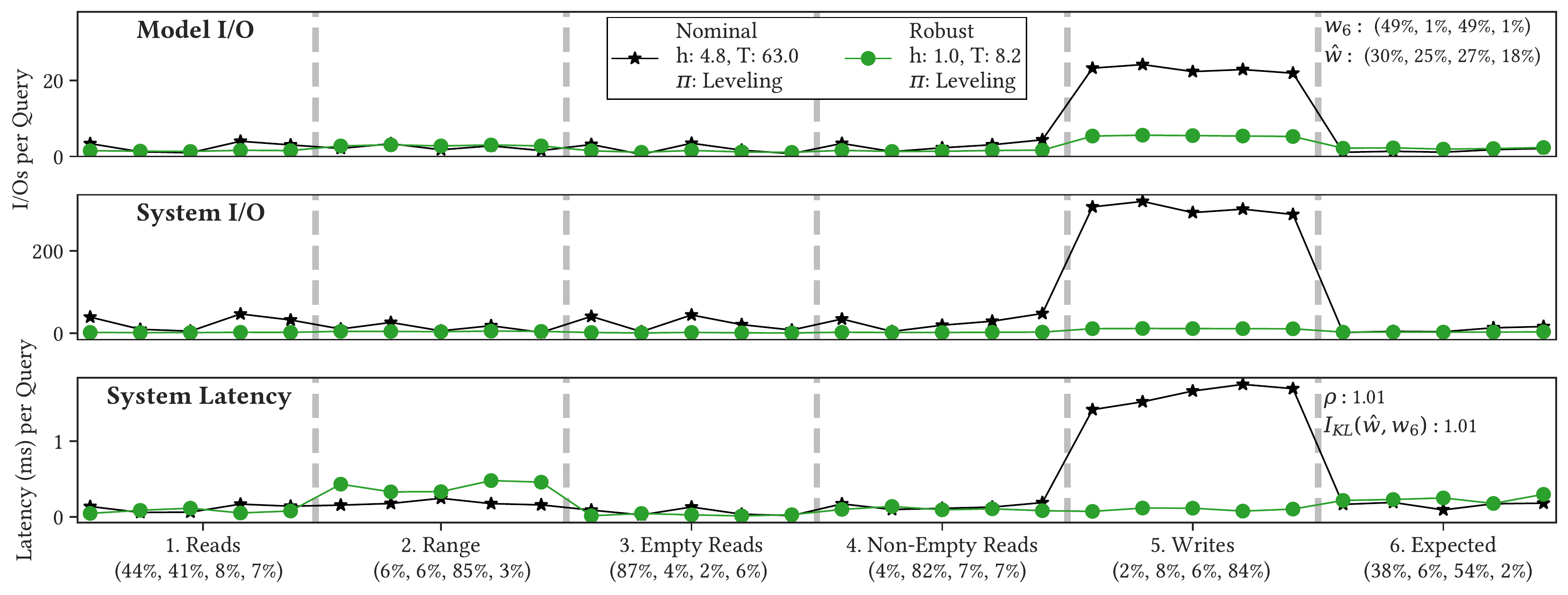}
    \caption{Bimodal workloads $w_{5}$ and $w_{6}$.}
    \label{fig:bimodal_wl_568}
\end{figure*}

\begin{figure*}
    \centering
    \includegraphics[width=\linewidth]{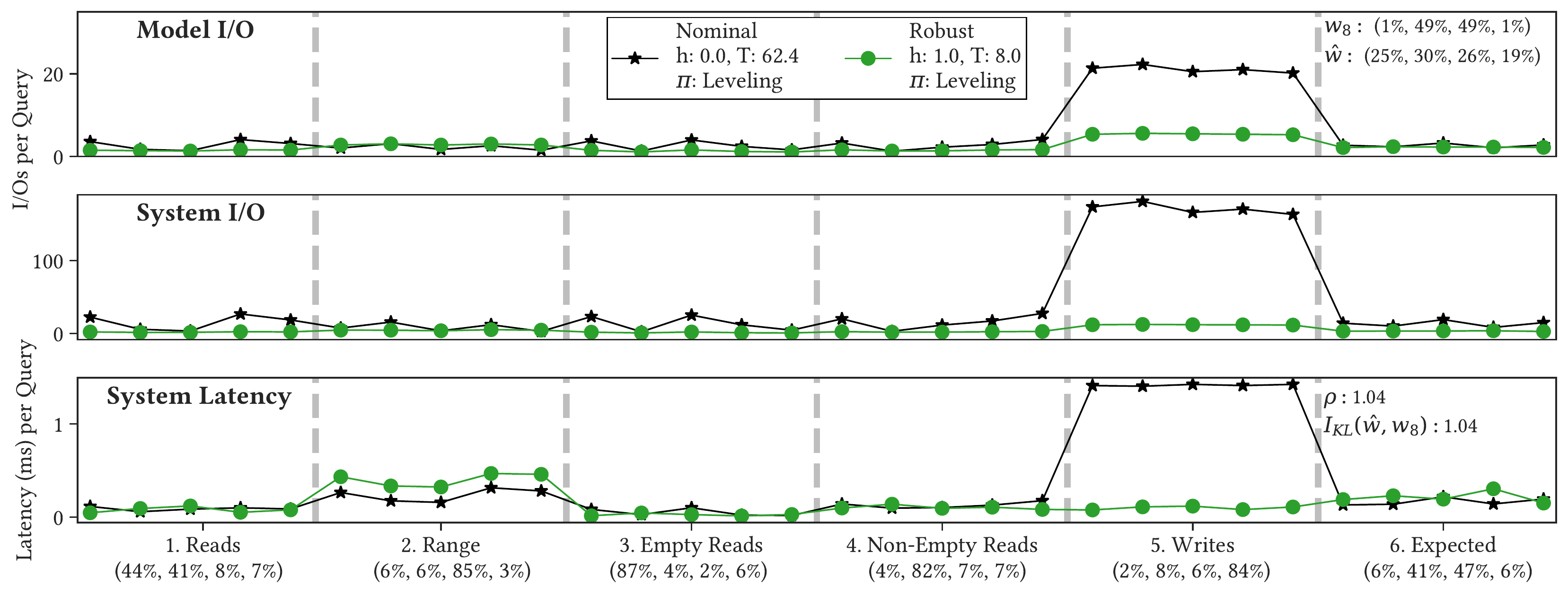}
    \includegraphics[width=\linewidth]{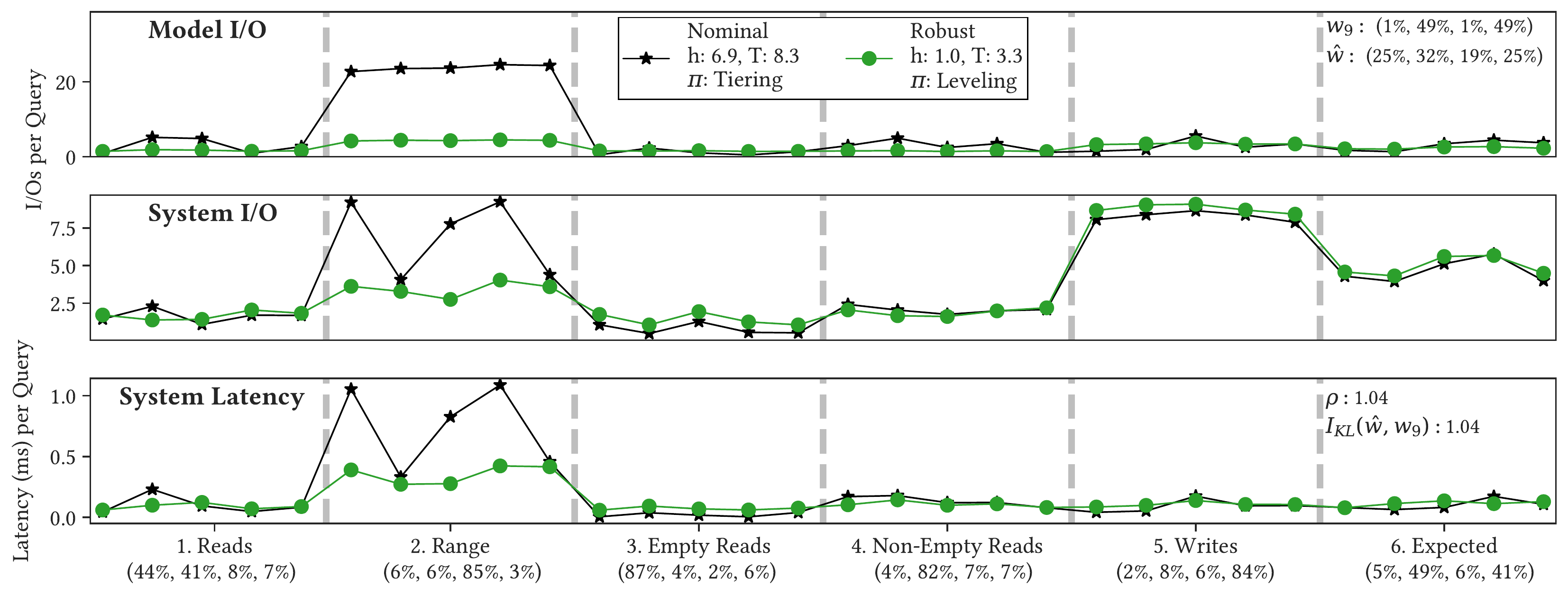}
    \includegraphics[width=\linewidth]{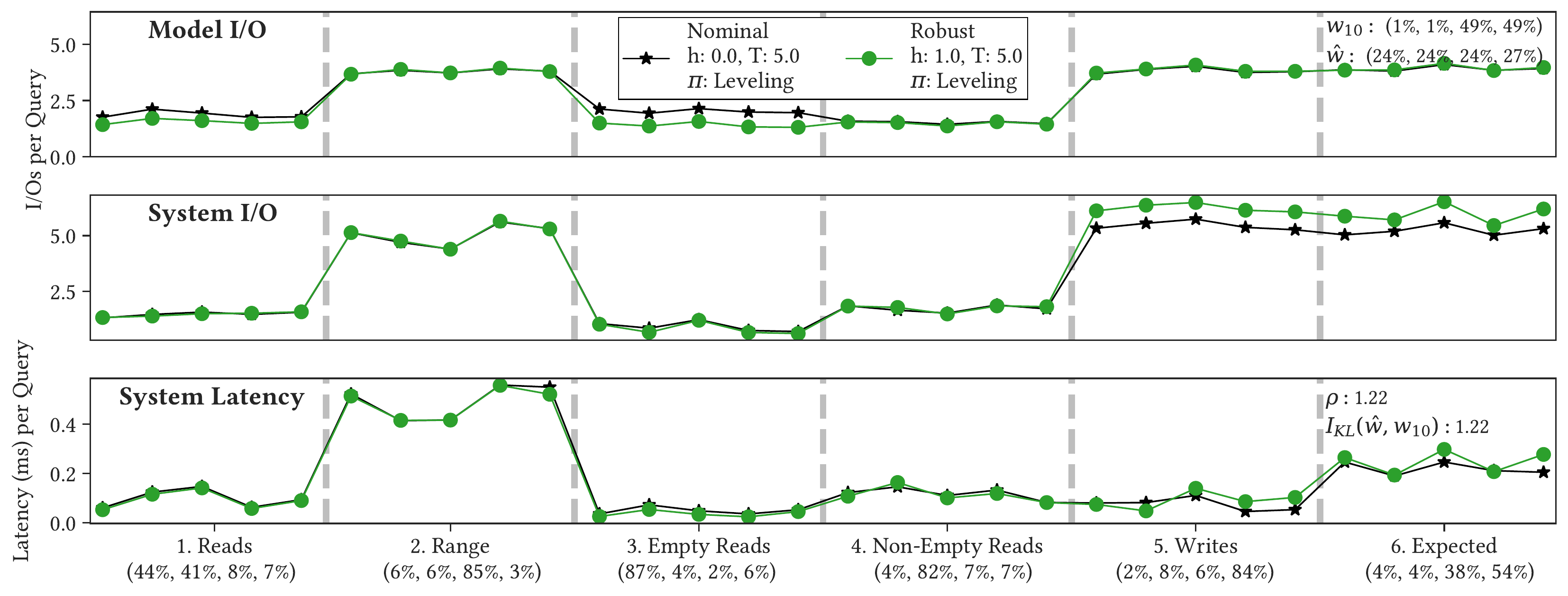}
    \caption{Bimodal workloads $w_{8}$, $w_{9}$, and $w_{10}$.}
    \label{fig:bimodal_wl_910}
\end{figure*}

\Paragraph{Biomodal Workloads}
In Figures~\ref{fig:bimodal_wl_568} and~\ref{fig:bimodal_wl_910} we see the
    remaining biomodal workloads.
For $\workload_{5}$ we see similar results as in $\workload_{1}$ and
    $\workload_{2}$ as these workloads are read dominated.
Tunings for $\workload_{6}$ and $\workload_{9}$ shows a similar difference in
    performance as $\workload_{3}$ as the expected workload contains a large
    fraction of range queries.
Both tunings for $\workload_{9}$ and $\workload_{10}$ show the system I/O
    jumping after the write session.
Again, this is due to the structure of the tree changing.
Note that $\workload_{9}$ in particular shows the mismatch between system and
    model I/O as our cost model does not account for the changes in tree
    structure.

\begin{figure*}
    \centering
    \includegraphics[width=\linewidth]{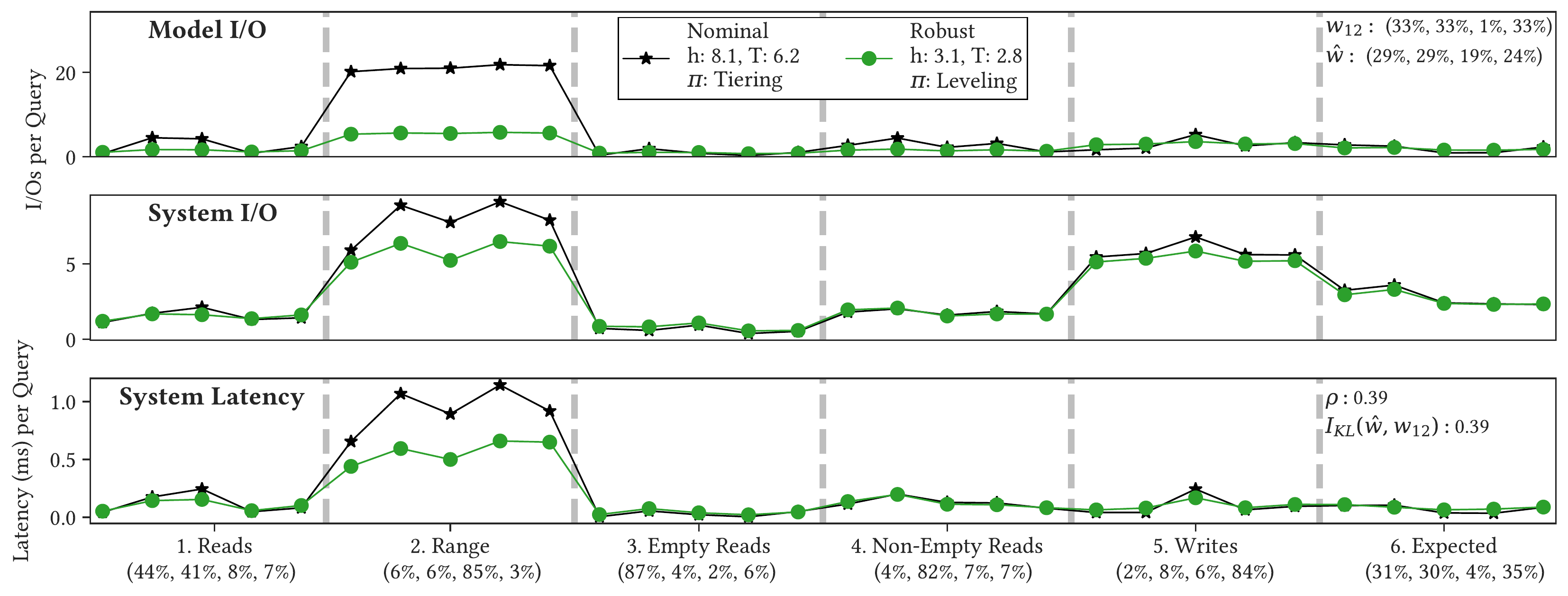}
    \includegraphics[width=\linewidth]{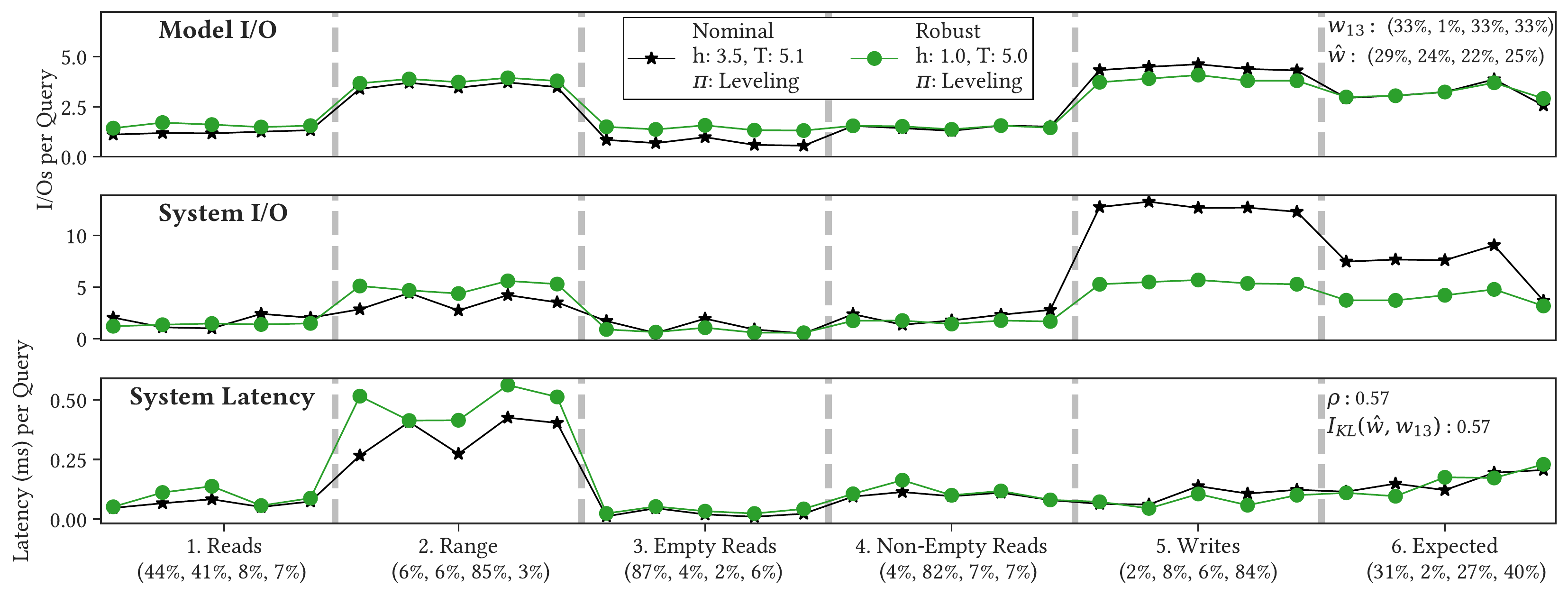}
    \includegraphics[width=\linewidth]{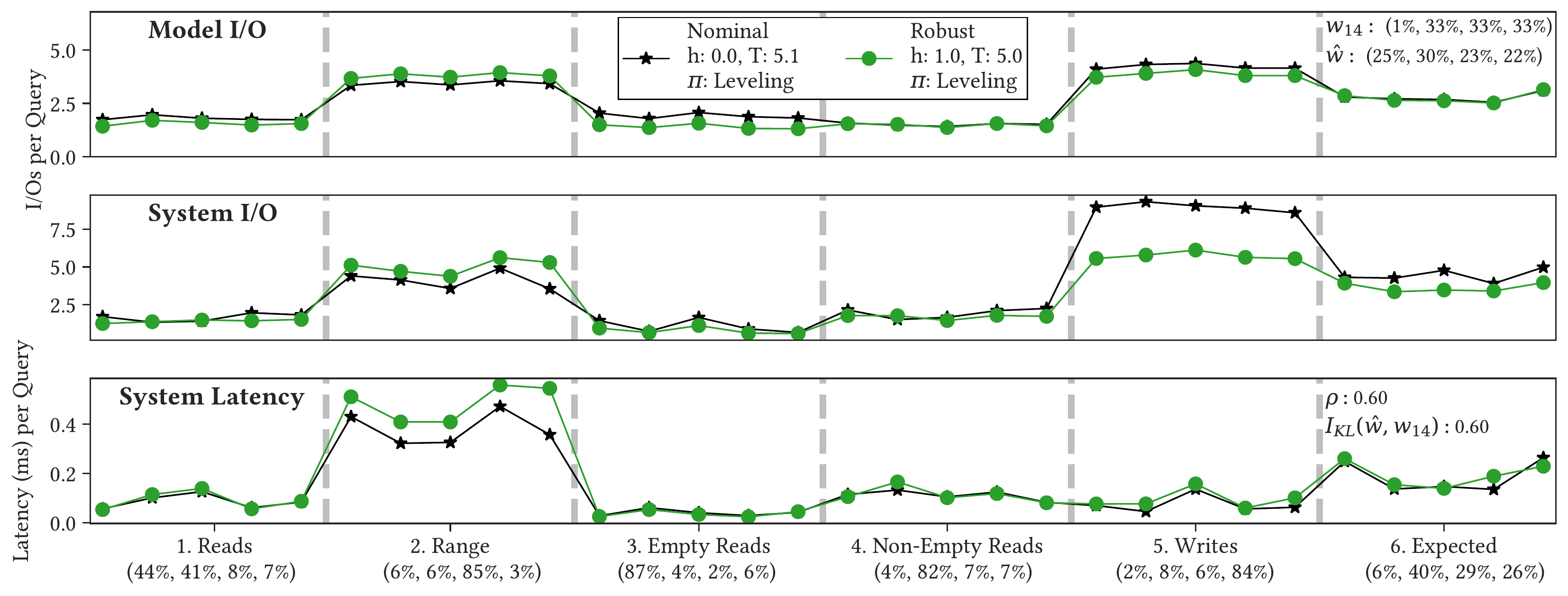}
    \caption{Trimodal workloads $w_{12}$, $w_{13}$, and $w_{14}$.}
    \label{fig:trimodal_wl}
\end{figure*}

\Paragraph{Trimodal Workloads}
In Figure~\ref{fig:trimodal_wl} we show tunings for the remaining trimodal
    workloads and their performance.
Because of the exclusion of range queries in the expected workload, both tunings
    for $\workload_{12}$ see an increase in system I/O and latency in the range
    query session.
However, the robust tuning does significantly better in both measurements as
    it lowers the bloom filter memory and size ratio.
Workloads $\workload_{13}$ and $\workload_{14}$ shows the robust tuning
    experiencing a trade-off of slightly worse range query performance compared to
    the nominal as evident in the range query session.
In return, the robust tuning causes the system to issue much less I/Os when a
    write heavy session occurs.
    
\Paragraph{Robust Outperforms Nominal for Properly Selected $\rho$}
In the model evaluation (Figure~\ref{fig:rho_vs_rho_hat}), we showed that robust
    tuning outperforms the nominal tuning in the presence of uncertainty for
    tuning parameter $\rho$ approximately greater than 0.2.
This is further supported by all the system experiments described.
We see the analysis of workload tunings above show instances where the
    KL-divergence of the observed workload averaged across all the sessions
    w.r.t. the expected workload is close to the tuning parameter $\rho$.
In each of these experiments, the robust tuning outperforms the nominal
    resulting in up to a $90\%$ reduction in latency and system I/O.
Conversely, when the observed workloads are similar to the expected one
    ($I_{KL}(\obsworkload, \workload_{11}) < 0.2$), such as in 
    Figure~\ref{fig:query_seq_reads_2}, we observe a resulting increase of
    latency by $20\%$.
    
\begin{figure}
    \centering
    \includegraphics[width=\linewidth]{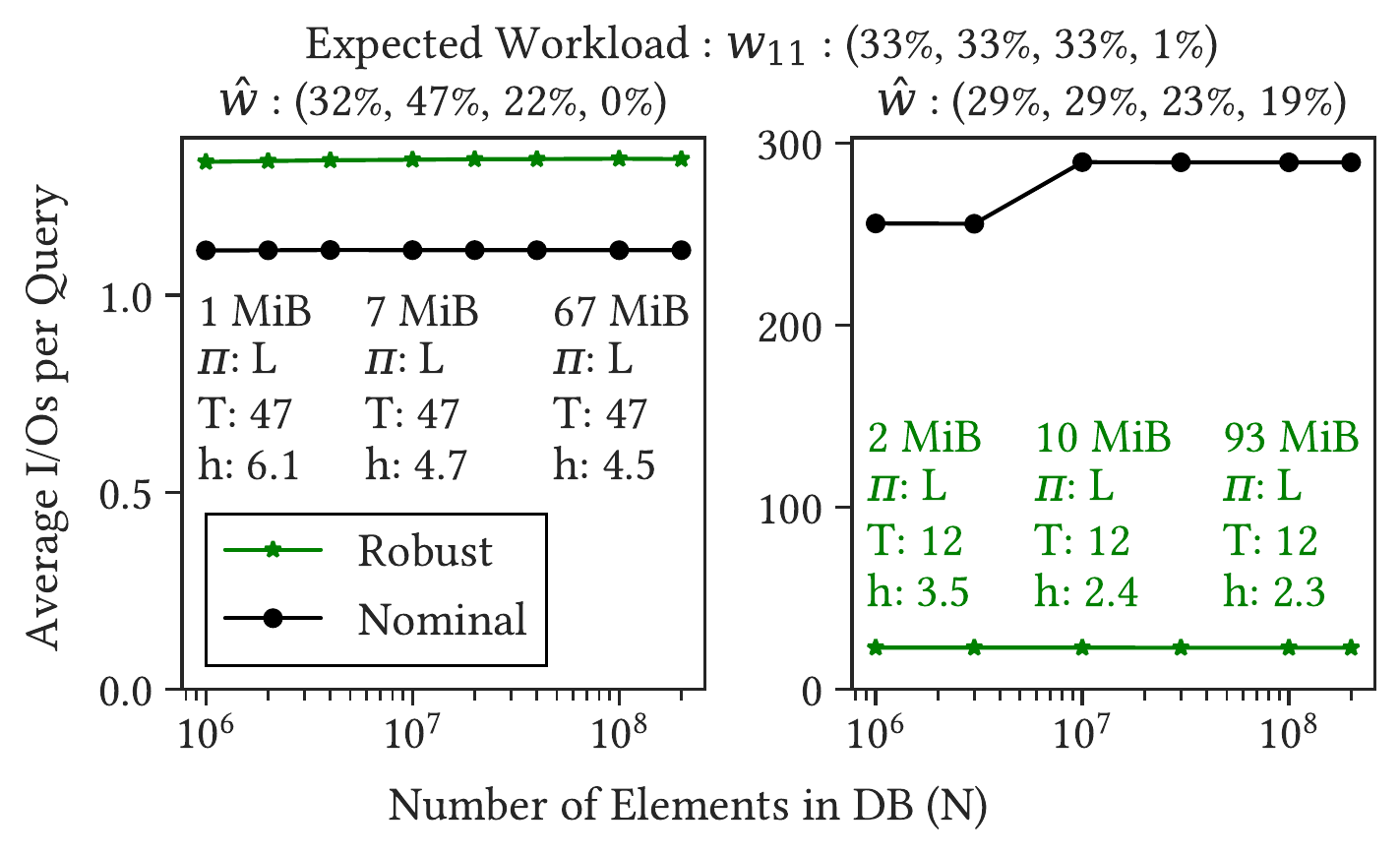}
    \caption{Impact of database size on performance. All tunings use the
        same expected workload $w_{11}$ with executed workloads shown above
        each graph. Points at each power of 10 show $m_{buf}$ and the tuning
        $\Phi$ (L for leveling, T for tiering).}
    \label{fig:perf_db_size}
\end{figure}

\Paragraph{ENDURE Scales with Data Size}
To verify that {\Endure} scales, we repeat the previous experiments, while
    varying the size of the initial database.
Each point in Figure~\ref{fig:perf_db_size} is calculated based on a series of
    workload sessions similar to the ones presented in
    Figures~\ref{fig:query_seq_reads_2} (\ref{fig:query_seq_hybrid_2}) for the
    left (right) part of Figure~\ref{fig:perf_db_size}.
All points use the same expected workload, therefore the nominal and
    robust tunings are the same across each graph.
We observe that the robust and nominal tuning increase buffer memory as the
    initial database size grows.
As a result, for all cases, the number of initial levels is the same regardless
    of the number of entries.
This highlights the importance of the number of levels w.r.t performance.
Additionally, the performance gap between robust and nominal stas consistent as
    database size grows, showing {\Endure} is effective regardless of data size.

\subsection{Robustness is All You Need}
One of the key challenges during the evaluation of tuning configurations in 
    presence of uncertainty is the challenge in measuring steady-state
    performance.
In Section~\ref{sec:system-results}, we show that the cost-model can accurately
    predict the empirical measurements.
In the course of this study, using our model, we compared over 700 different
    robust tunings with their nominal counterparts over the uncertainty
    benchmark set {\benchmark}, leading to approximately 8.6 million
    comparisons.
Robust tunings comprehensively outperform the nominal tunings in over 80\% of
    these comparisons.
We further cross-validated the relative performance of the nominal and the 
    robust tunings in over 300 comparisons using RocksDB.
The empirical measurements overwhelmingly confirmed the validity of our 
    analytical models, and the few instances of discrepancy in the scale of 
    measured I/Os, such as the ones discussed in previous sections, are 
    easily explained based on the structure of the LSM tree.

\Paragraph{Leveling is ``more'' Robust than Tiering}
One of the key takeaways of applying robust tuning to LSM trees is that
    \emph{leveling is inherently more robust} to perturbations in
    workloads when compared to pure tiering.
This observation is in line with the industry practice of deploying leveling or
    hybrid leveling over pure tiering.
Overall, based on our analytical and empirical results, the robust tuning should
    always be employed when tuning an LSM tree, unless the future workload
    distribution is known with absolute certainty.

\Paragraph{Discussion}
While we have deployed and tested robust tuning on LSM trees, the robust
    paradigm of {\Endure} is a generalization of a minimization problem that is
    at the heart of any database tuning problem.
Hence, similar robust optimization approaches can be applied to \emph{any
    database tuning} problem assuming that the underlying cost-model is known,
    and each cost-model component is convex or can be accurately approximated by
    a convex surrogate.

\balance
\section{Related Work}
\label{sec:related_work}

\Paragraph{Tuning Data Systems}
Database systems are notorious for having numerous tuning knobs. 
These tuning knobs control fine-grained decisions (e.g., number of threads, 
    amount of memory for bufferpool, storage size for logging) as well as basic 
    architectural and physical design decisions about partitioning, index 
    design, materialized views that affect storage and access patterns, and 
    query execution \cite{Bruno2005,Chaudhuri1998a}. 
The database research community has developed several tools to deal with such 
    tuning problems. 
These tools can be broadly classified as offline workload analysis for index 
    and views design 
    \cite{Agrawal2004,Agrawal2000,Chaudhuri1997,Dageville2004,Valentin2000,Zilio2004},
    and periodic online workload analysis 
    \cite{Bruno2006,Schnaitter2006,Schnaitter2007,Schnaitter2012} to capture
    workload drift \cite{Holze2010}. 
In addition, there has been research on reducing the magnitude of the search 
    space of tuning \cite{Bruno2005,Dash2011} and on deciding the optional data 
    partitioning 
    \cite{Athanassoulis2019,Papadomanolakis2004,Serafini2016,Sun2014,Sun2016}.
These approaches assume that the input information about resources and workload 
    is accurate. 
When it is proved to be inaccurate, performance is typically severely impacted.

\Paragraph{Adaptive \& Self-designing Data Systems}
A first attempt to address this problem was the design of \textit{adaptive} 
    systems which had to pay additional transition costs
    (e.g., when deploying a new tuning) to accommodate shifting workloads 
    \cite{Idreos2007,Graefe2010a,Graefe2010c,Schuhknecht2018}. 
More recently the research community has focused on using machine learning to 
    learn the interplay of tuning knobs, and especially of the knobs that are 
    hard to analytically model to perform cost-based optimization. 
This recent work on self-driving database systems 
    \cite{Aken2017,Ma2018,Pavlo2017} or self-designing database systems 
    \cite{Idreos2019,Idreos2019a,Idreos2018a,Idreos2018} 
    is exploiting new advancements in machine learning to tune database systems 
    and reduce the need for human intervention, however, they also yield 
    suboptimal results when the workload and resource availability information 
    is inaccurate. 

\Paragraph{Robust Database Physical Design}
One of the key database tuning decisions is physical design, that is, the 
    decision of which set of auxiliary structures should be used to allow for 
    the fastest execution of future queries. 
Most of the existing systems use past workload information as a representative 
    sample for future workloads, which often leads to sub-optimal decisions when 
    there is significant workload drift. 
Cliffguard \cite{Mozafari2015} is the first attempt to use unconstrained robust 
    optimization to find a robust physical design. 
Their method is derived from Bertsimas et al. in~\cite{Bertsimas2010-qo}, a
    numerical optimization approach using alternating gradient ascent-descent to
    optimize problems without closed-form objectives.
In contrast to Cliffguard, {\Endure} focuses on the LSM tree tuning problem
    which uses an analytical closed form objective in Equation~\eqref{eq:thecost}.
This allows us to directly solve a Lagrangian dual problem instead of relying
    upon numerical optimization techniques.
Furthermore, we found that the approach in Cliffguard, when applied to our
    objective, fails to converge even after extensive hyperparameter search.
    

\section{Conclusion}
\label{sec:conclusion}

In this work, we explored the impact of workload uncertainty on the performance
    of LSM tree-based databases, and introduced {\Endure}, a robust tuning
    paradigm that recommends optimal designs to mitigate any performance
    degradation in the presence of workload uncertainty.
We showed that in the presence of uncertainty, robust tunings increase database
    throughput compared to standard tunings by up to $5 \times$.
Additionally, we provided evidence that our analytical model closely matches the
    behavior measured on a database system.
Through both model-based and extensive experimental evaluation of {\Endure} 
within the state-of-the-art LSM-based storage engine, RocksDB, we show that the robust 
tuning methodology consistently outperforms classical tuning strategies. 
{\Endure} can be used as an indispensable tool for database administrators
    to both evaluate the performance of tunings, and recommend optimal tunings
    in a few seconds without resorting to expensive database experiments.

\clearpage
\bibliographystyle{ACM-Reference-Format}
\bibliography{bibliography,library}
\balance
\end{document}